\documentclass[%
 reprint,
nofootinbib,
 amsmath,amssymb,
 aps,
]{revtex4-2}
\usepackage{graphicx}
\usepackage{dcolumn}
\usepackage{bm}
\usepackage[hidelinks]{hyperref}
\usepackage{color}
\usepackage{physics}
\usepackage{dsfont}
\usepackage{pgf,tikz}
\usepackage{pgfplots}
\usepackage{adjustbox}
\usepackage{soul}
\usepackage{amsmath,amssymb,physics}
\usepackage{rotating} 
\usepackage{comment}
\allowdisplaybreaks[3]

\hypersetup{citecolor=red,colorlinks=true,urlcolor=blue}

\usepackage{lmodern}
\usepackage[T1]{fontenc}

\newcommand{\IOPPAS}{Institute of Physics PAS, Aleja Lotnik\'ow 32/46, 02-668 Warszawa, Poland}

\newcommand{\Sarrow}[3]{\hat S^{#1\rightarrow #2}_{#3}}
\DeclareMathOperator{\sgn}{sgn}

\begin{document}

\preprint{APS/123-QED}

\title{
Population eigenstates of the SU(d) spin-exchange model for high-spin fermions in optical lattices
}

\author{Hubert Dunikowski and Emilia Witkowska}
\affiliation{\IOPPAS}

\begin{abstract}
We investigate the $\mathrm{SU}(d)$ spin exchange model describing ultra-cold fermionic atoms with spin $s\ge 1$ in a one-dimensional optical lattice. The model emerges from the Fermi-Hubbard model in the strongly interacting regime with one atom in each lattice site.
The central result of this work is the systematic construction of eigenstates in terms of magnetic sub-level populations, which we call population eigenstates.
Exploiting this framework, we derive effective light-induced Hamiltonians via a second-order Schrieffer–Wolff transformation projected onto the population eigenstates. The resulting models reveal a qualitative difference between spin-1/2 and higher-spin systems: whereas spin-1/2 dynamics remains confined to the maximal-spin Dicke manifold, the extensive $\mathrm{SU}(d)$ degeneracies for $s\ge 1$ allow coherent population transfer across sectors of different collective spin length, generating unconventional spin dynamics that cannot be captured by any fixed-spin-manifold description. Agreement with exact Fermi-Hubbard dynamics confirms the framework as a practical foundation for quantum-enhanced correlations and metrological protocols in high-spin fermionic systems.
\end{abstract}

\date{\today}

\maketitle

\section{Introduction}

The $\mathrm{SU}(d)$ symmetry of alkaline-earth and alkaline-earth-like fermionic atoms in optical lattices gives rise to an extraordinarily rich many-body landscape, characterized by extensive degeneracies and collective spin dynamics that have no counterpart in conventional spin-1/2 systems. This symmetry emerges from the decoupling of electronic and nuclear spin $s$ degrees of freedom giving rise to highly symmetric models with $d=2s+1$ magnetic sub-levels \cite{Gorshkov2010,Cazalilla2014}, enabling the exploration of many-body physics beyond the conventional spin-$1/2$ paradigm \cite{Taie2012,Scazza2014,Pagano2014,Zhang2014}. Experimental implementations with isotopes such as $^{87}$Sr and $^{173}$Yb have already demonstrated long-lived quantum magnetism, spin-exchange interactions, and collective coherent dynamics in systems possessing up to ten internal spin components \cite{Cappellini2014,Bromley2018}.

The richness of $\mathrm{SU}(d)$-symmetric systems is most clearly exposed in the strongly interacting limit of the Fermi–Hubbard model with one atom in each lattice site, where dynamics is governed by spin-exchange processes. 
For spin-1/2 fermions this limit yields the isotropic Heisenberg model \cite{Auerbach1994}, whose eigenstates carry a unique, well-defined collective spin value for each energy sector. The situation is fundamentally different for $s\ge 1$.
In this case, because of a simple permutation-like form of the dynamics, the effective Hamiltonian is described by the $\mathrm{SU}(d)$ spin exchange (SE) model with $d>2$, whose symmetry and degeneracy structure become increasingly rich with growing $d$ \cite{Gorshkov2010,Cazalilla2014}. In particular, within a given energy sector of the SE model, various values of collective spin are allowed, in contrast to spin-$1/2$ systems.
While the symmetry properties of $\mathrm{SU}(d)$ magnets have been extensively studied \cite{PhysRevLett.54.966,Gorshkov2010}, a complete and physically transparent characterization of the SE eigenstates that simultaneously exploits these degeneracies and connects directly to experimentally accessible observables has remained missing.

In this work, we provide such a framework by systematically constructing the set of eigenstates of the $\mathrm{SU}(d)$ SE Hamiltonian in terms of population eigenstates (PES).
The construction is rooted in a key physical observation: the SE Hamiltonian conserves the global population of each magnetic sub-level individually, since spin-exchange processes only swap the internal states of atoms between neighboring sites without creating or destroying population in any sub-level.
As a consequence, the Hilbert space naturally decomposes into sectors of fixed population configuration rather than by collective spin values, a representation directly accessible in modern cold-atom experiments where magnetic sub-level populations are measured with single-atom resolution~\cite{ferrer2026,plassmann2026}.

We identify two physically relevant classes of PES within the energy manifolds governing the strongly interacting regime. The first, ground population eigenstates (GPES), span the degenerate zero-energy manifold; a non-trivial result of our construction is that each magnetic sub-levels populations uniquely labels a single GPES.
The second class, $q$-wave population eigenstates ($q$-PES), forms the neighboring energy manifolds; generated through collective spin-transfer operators carrying a well-defined quasi-momentum $q$.
Building on the PES decomposition, we further construct the collective spin eigenstates to determine the allowed values of the collective spin within each energy manifold, revealing that for $s\ge 1$ a broad range of collective spin lengths coexists within each energy sectors,  in stark contrast to the spin-1/2 case where each energy sector supports a well defined collective spin value.

The utility of the PES framework is illustrated by deriving effective light-induced Hamiltonians for high-spin fermions subject to spin-orbit coupling. Using a Schrieffer--Wolff transformation~\cite{BRAVYI20112793} with projection onto the PES manifolds, and thereby properly accounting for the full degeneracy structure of the SE model, we obtain effective models that systematically capture the physics inaccessible to descriptions restricted to a single collective-spin manifold. The resulting Hamiltonian accurately reproduces the dynamics of the full Fermi--Hubbard model with weak light coupling over experimentally relevant timescales. 
We characterize the total spin dynamics, and show that the light coupling drives coherent population transfer between sectors of different collective spin length, a mechanism that produces pronounced oscillations of total spin and qualitatively modifies the generation of spin-spin correlations compared to what a fixed-spin-manifold description would predict.The PES framework therefore provides both a complete eigenstate characterization of the considered energy manifolds of the $\mathrm{SU}(d)$ SE model and a practical foundation for deriving controlled effective theories in high-spin fermionic systems.

The paper is organized as follows. In Sec.~II, we introduce the Fermi--Hubbard and SE models. Section~III presents the construction of the GPES and the $q$-PES. In Sec.~IV, we construct the spin eigenstates of the SE model and determine the collective spin values permitted within each energy manifold. Section~V derives the effective light-induced Hamiltonians and validates their accuracy against the full microscopic model. Finally, Sec.~VI summarizes our findings and discusses their implications.

\section{The model}

\subsection{The Fermi-Hubbard model}

We describe the system of $N$ fermionic atoms with spin $s$ in a one-dimensional optical lattice by the single-band Fermi-Hubbard Hamiltonian~\cite{Cazalilla2014}, which assumes spin-independent contact interactions and neglects higher-band contributions. The atoms are described in the Wannier basis, where each atom occupies one of $d=2s+1$ internal magnetic sub-levels. The state of the single atom can be described by two integers: the site index $j$ (from 1 to $N$) and the magnetic quantum number $m$ (from $-s$ to $s$).

The Hamiltonian is composed of the spin-preserving tunneling term of atoms among neighbouring sites ($\hat{H}_{\mathrm{t}}$ governed by $\mathrm{J}$), and spin independent on-site repulsion ($\hat{H}_{\mathrm{int}}$ governed by $U$). This gives rise to the Fermi-Hubbard Hamiltonian
\begin{equation}
\label{eq:HFH}
    \hat{H}_{\mathrm{FH}} = \hat{H}_{\mathrm{t}} + \hat{H}_{\mathrm{int}} ,    
\end{equation}
\begin{align}
    \label{eq:FHtunneling}
    \hat{H}_{\mathrm{t}} &= -\mathrm{J}\sum\limits_{j}\sum\limits_{m=-s}^{s} (\hat a_{j+1,m}^\dagger \hat a_{j,m} + \text{h.c.}) ,\\
    \label{eq:FHint}
    \hat{H}_{\mathrm{int }} &= \frac{U}{2}\sum\limits_{j=1}^{N} \hat n_j(\hat n_j-1)\, ,
\end{align}
where $\hat a_{j,m}$ ($\hat a^\dagger_{j,m}$ ) is an annihilation (creation) operator for atom in the $j$-th site and the $m$th magnetic sub-level, and where $\hat{n}_j$ is a particle number operator for a $j$-th site: $\hat{n}_j=\sum_{m=-s}^s\hat a_{j,m}^\dagger \hat a_{j,m}$.
Throughout this work we assume periodic boundary conditions, so that site $j = N+1$ is identified with site $j = 1$.

\subsection{The Spin Exchange model}

In the regime when repulsion strength is larger than tunneling ($U\gg\mathrm{J}$), with the number of atoms in the system $N=\sum_{j=1}^N n_j$ equal to the number of sites of the lattice, the spectrum of FH Hamiltonian reveals a band structure.
The Hamiltonian $\hat{H}_{\mathrm{FH}}$ is then dominated by $\hat{H}_{\mathrm{int}}$ responsible for emergence of bands spaced by $U$ (different bands correspond to states with different multi-atomic occupations of single sites) while $\hat{H}_{\mathrm{t}}$ coupling between these bands can be treated as a perturbation.
The lowest-energy manifold consists of states with exactly one atom per lattice site (singlons). Since these states contain no doubly occupied sites, they are unaffected by contact interactions and therefore have zero interaction energy.

\begin{figure}[]
    \Large \raisebox{70pt}{(a)}\ \includegraphics[width=0.5\linewidth]{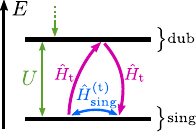}\\[15pt]
    \Large \raisebox{50pt}{(b)}\ \includegraphics[width=0.9\linewidth]{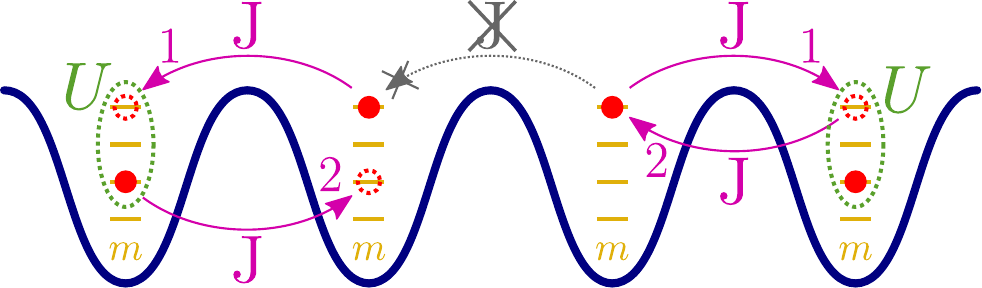}\\[15pt]
    \Large \raisebox{50pt}{(c)}\ \includegraphics[width=0.9\linewidth]{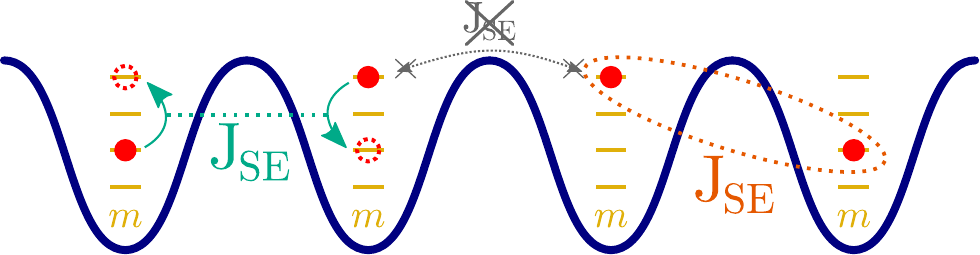}
    \caption{
    Schematic representation of the second-order perturbative process yielding the effective spin-exchange Hamiltonian $\hat{H}_{\rm SE}$ from the Fermi-Hubbard model in the strongly interacting regime ($U\gg {\rm J}$) using \eqref{eq:pert_FH_to_SE}. 
    (a) Energy diagram showing the virtual excitation from the singlon manifold to the doublon band and back, with energy denominator $U$ setting the scale ${\rm J}_{\rm SE} = 2{\rm J}^2/N$. 
    (b) The corresponding virtual processes on the lattice: an atom tunnels to a neighboring site (creating a doublon), then returns, effectively exchanging the spin states of the two sites. (c) The emergent effective nearest-neighbor spin-exchange processes described by $\hat{H}_{\mathrm{sing}}^{(\mathrm{t})}=\hat{H}_{\mathrm{SE}}$.}
    \label{fig:fig1}
\end{figure}

In the strongly interacting regime ($U\gg\mathrm{J}$), interband transitions are strongly suppressed, such that a state initially prepared in a given band evolves predominantly within that band. 
For the singlon manifold, however, the projected Fermi-Hubbard Hamiltonian is trivial,
$\hat P(\text{\small sing})\hat{H}_{\mathrm{FH}} \hat P(\text{\small sing})=0$, and therefore does not generate any dynamics. 
The leading nontrivial contribution arises at second order in the tunneling Hamiltonian through virtual excitations to the dublon manifold  (vide Fig.~\ref{fig:fig1}):
\begin{equation}
    \hat{H}_{\text{sing}}^{(\mathrm{t})} = \hat P(\text{\small sing})\,\hat{H}_{\mathrm{t}}\,\frac{\hat P(\text{\small dub})}{E_{\text{sing}}-E_{\text{dub}}}\,\hat{H}_{\mathrm{t}}\,\hat P(\text{\small sing})\label{eq:pert_FH_to_SE}
\end{equation}
where $(\hat P_{\rm sing})$ and $(\hat P_{\rm dub})$ project onto the singlon and dublon manifolds, respectively. Here, the dublon manifold denotes the lowest excited band containing a single doubly occupied site. The unperturbed energies are $E_{\text{sing}}=0$ and $E_{\text{dub}}=U$.

The second order perturbation $\hat{H}_{\text{sing}}^{(\mathrm{t})} $ \eqref{eq:pert_FH_to_SE} yields the spin-exchange Hamiltonian $\hat{H}_\text{SE}$:
\begin{equation}
\hat{H}_\text{SE} =
\underbrace{\frac{2\mathrm{J}^2}{U}}_{\mathrm{J_{SE}}}
\sum\limits_{j=1}^{N}
\sum\limits_{m\neq m'} 
\Big(
\hat{S}^{m'\rightarrow m}_j \hat{S}^{m\rightarrow m'}_{j +1}
-\hat{n}_{j,m} \hat{n}_{j+1,m'}
\Big),
\label{eq:HSE}
\end{equation}
where
\begin{align}
    \hat{n}_{j,m} & = \hat{a}^\dagger _{j, m} \hat{a} _{j, m}\\
    \hat{S}^{m\rightarrow m'}_j & =
    \hat{a}^\dagger _{j, m'} \hat{a} _{j, m}
\end{align}
$\hat n_{j,m}$ are the number operator of atoms in the $j$-th lattice site and $m$ magnetic quantum number, and the S-arrow operator $\hat{S}^{m\rightarrow m'}_j$ is the spin changing operator between the two magnetic sub-levels from $m$ to $m'$ at site $j$.
These operators are the fundamental building blocks of the SE Hamiltonian and will play a central role throughout this work; their algebraic properties are collected in Appendix \ref{apx:S-arrow}.

The SE model consists of two parts, corresponding to two physical processes. One is an emergent nearest neighbour interaction coming from the process when the atom tunnels to the neighbouring site to feel the on-site repulsion and return back to its site of origin. The other is an emergent exchange of the spins between neighbouring sites (hance the name of the Hamiltonian) coming from a physical exchange of the atoms between neighbouring sites. It is important to notice, that these processes vanish when neighboring atoms occupy the same magnetic sub-level, since the Pauli exclusion principle forbids tunneling into an already occupied state of the same spin (this is reflected by the $m\neq m'$ exclusion in the second summation sign in \eqref{eq:HSE}).

The form of the SE Hamiltonian presented in \eqref{eq:HSE} is the most natural one with terms directly corresponding to the physical processes behind the effective dynamics. It is however possible to rewrite the SE Hamiltonian in numerous forms seemingly different, yet in fact equivalent to the one presented in \eqref{eq:HSE}, see Appendix \ref{apx:alt_SE} for details.

The on-site spin operators $\hat{J}_{\sigma,j}$, can be expressed in terms of S-arrow operators as follows:
\begin{align}
\hat{J}_{z,j} &= \sum_{m=-s}^{s}\! m\ \hat{S}_j^{m\rightarrow m} ,\\
\hat{J}_{+,j} &= \sum_{m=-s}^{s-1}\! \alpha_{s,m}
\hat{S}_j^{m\rightarrow m+1} ,\\
\hat{J}_{-,j} &= \sum_{m=-s}^{s-1}\! \alpha_{s,m}\ \hat{S}_j^{m+1\rightarrow m} ,
\end{align}
where
\begin{equation}
\alpha_{s,m} = \sqrt{(s-m)(s+m+1)},\label{eq:alpha}
\end{equation}
and 
$\hat{J}_{x,j } = (\hat{J}_{+,j}+\hat{J}_{-,j})/2$, $\hat{J}_{y,j} = (\hat{J}_{+,j}-\hat{J}_{-,j})/(2i)$.
The collective components of spin operators are
$\hat{J}_\sigma = \sum_j \hat{J}_{\sigma,j}$.
Therefore, the total spin $\hat{J}^2 = \hat{J}^2_{x} + \hat{J}^2_y + \hat{J}^2_z$ defines the spin states
\begin{align}
\hat{J}^2\,|\Lambda,M\rangle &= \Lambda(\Lambda+1)\,
|\Lambda,M\rangle , \\
\hat{J}_z\,|\Lambda,M\rangle &= M\,|\Lambda,M\rangle.
\end{align}

It is worth noticing here that for $s\geq 1$ the S-arrow operators can not be identified with the spin operators, and hence, the spin states can not be identified with the eigenstates of the collective S-arrow operators, as it is true for $s=\frac{1}{2}$.

It is also convenient to define the collective S-arrow operators:
\begin{equation}
    \hat{S}^{m\rightarrow m'} = \sum_{j=1}^N \hat{S}^{m\rightarrow m'}_j.
\end{equation}
They bring an atom from $m$ to $m'$ magnetization no matter on which sit it seats, so they operate on global populations of atomic magnetization $\hat n_m=\sum_{j=1}^N \hat n_{j,m}$ rather than on a on-site populations.
In particular, the diagonal collective S-arrow operators $\hat{S}^{m\rightarrow m}$ coincide with the collective on-site population operators of magnetic sub-level $m$, a property that will be central to the construction of population eigenstates in Sec.~\ref{sec:PES}.

\section{PES: Population Eigenstates of $\hat H_{\textrm SE}$}
\label{sec:PES}

The spin-exchange Hamiltonian \eqref{eq:HSE} possesses a rich conserved-quantity structure; it commutes with both the collective S-arrow operators and the collective spin operators
\begin{align}
    \big[\hat{H}_\text{SE}, \hat{S}^{m\rightarrow m'} \big] &= 0\label{eq:HSE_S-arrow_kom}\\[5pt]
    \Big[\hat{H}_\text{SE}, \hat{\vec J} \Big] &= 0 ,
\end{align}
therefore, its eigenbasis can be written in the form common with S-arrow and the collective spin $\hat J^2$ operators (there is some degeneracy).
Since $\hat H_{\text{SE}}$ commutes with the collective S-arrows, it preserves populations of atomic magnetization levels, because such a population is equal to a specific collective S-arrow $ \hat{S}^{m\rightarrow m} = \sum_{j=1}^N \hat{S}^{m\rightarrow m}_j=\hat n_m$. Therefore, it is expected that SE Hamiltonian posses eigenstates, that have specific overall populations of atomic magnetic sub-levels $n_m$. 
We denote the population configuration of an eigenstate by $\vec{n}=(n_s,n_{s-1},\ldots,n_{-s})$, where $n_m$ is the population of the magnetic sub-level $m$.
The set of this populations $\vec{n}$ provide an additional set of quantum numbers, allowing the eigenstates to be labeled as $|E,\vec{n}\rangle$. We refer to such states as \emph{population eigenstates} (PESs).

In the context of solid state systems PES states would be rather unpractical and hard to encounter, since in experiments for these systems the observed quantity is a global magnetic response of the sample, hence the observed states are rather collective spin states. However in the context of modern cold-atom lattice experiments~\cite{ferrer2026, plassmann2026}, where populations of specific single-atom magnetic sub-levels are directly measured, the PES states comes as a very natural ones.

Although $\hat H_{\mathrm{SE}}$ commutes with both the S-arrow and collective-spin operators, the S-arrow operators generally do \emph{not} commute with the collective-spin operators.
The collective S-arrow operators do not generally commute among themselves \eqref{eq:S-qrrow_col_kom}. An important exception is provided by the diagonal operators $\hat{S}^{m\rightarrow m} =n_m$, which correspond to magnetic sub-level populations and mutually commute. Consequently, one may construct a common eigenbasis of $\hat H_{\mathrm{SE}}$ and the population operators $\hat n_m$ (PES base), or a common eigenbasis of $\hat H_{\mathrm{SE}}$ and the total spin operators ($|\Lambda,M\rangle$ base).
However, no basis can simultaneously diagonalize $\hat H_{\mathrm{SE}}$, all $\hat n_m$ and collective spin operators. As a result, PESs are generally not eigenstates of the collective spin.

Although PESs are generally not eigenstates of the collective spin, they remain eigenstates of the collective magnetization operator $\hat J_z$: $\hat J_z\,|E,\vec{n}\rangle=\big(\sum_{m=-s}^s m\,n_m\big)\,|E,\vec{n}\rangle$.
The corresponding eigenvalue is determined solely by the populations of the magnetic sub-levels.

\subsection{Ground population eigenstates (GPESs)}

The SE manifold of zero energy $E=0$ can be spanned by \emph{Ground Population Eigenstates (GPES)}.\footnote{These states are ground states for $\mathrm{J_{SE}}<0$ and actually the highest energy states for $\mathrm{J_{SE}}>0$.}
All GPESs have zero energy, $E=0$, and are uniquely specified by the set of all collective populations of magnetic sub-levels $\vec{n}=(n_s,n_{s-1},\ldots,n_{-s})$; that is, no two distinct GPESs share the same population configuration. We therefore omit the energy label and denote these states simply by $|\vec{n}\rangle$:
\begin{equation}
    \hat H_\mathrm{SE} \,|\vec{n}\rangle = 0\,|\vec{n}\rangle
\end{equation}
here $n_m=\sum_j n_{j,m}$, where note that we have $n_{j,m}=0,1$.

The simplest GPES are \emph{the sea states} $|\mu\rangle$, the ones that have all atoms occupying the same magnetization sub-level $m=\mu$:
\begin{equation}
   |\mu\rangle\equiv \Big|\vec{n}\!=\!(0,\ldots,\underbrace{N}_{n_\mu},\ldots,0)\Big\rangle  =\bigotimes\limits_{j=1}^N |\mu\rangle_j.
\end{equation}
Here, $|\mu\rangle_j$ denotes the state of the atom at site $j$ in the magnetic sub-level $m=\mu$. These states are eigenstates of $\hat H_{\mathrm{SE}}$ with $E=0$. Indeed, because all atoms occupy the same spin state, Pauli exclusion forbids spin-exchange processes, causing every term in ($\hat H_{\mathrm{SE}}$) to annihilate ($|\mu\rangle$).

As the spin-exchange Hamiltonian commutes with the collective S-arrow operators, $\hat{S}^{m\rightarrow m'}$ \eqref{eq:HSE_S-arrow_kom}, all GPESs can be generated from an arbitrary sea state $|\mu\rangle$ through successive applications of the appropriate S-arrows operators,
\begin{equation}
\label{eq:PES}
    |\vec{n}\rangle 
    = \mathcal{N}_{\vec{n},\mu}^{-1/2}
    \prod_{m\in\vec{m}_\mu} \left( \hat{S}^{\mu \rightarrow m} \right)^{n_m} |\mu \rangle\, ,
\end{equation}
where $\vec{m}_\mu$ denotes the set of magnetic sub-levels excluding $\mu$:
\begin{equation}
    \vec{m}_\mu=(s,\ldots,{\mu+1},{\mu-1},\ldots,{-s}).
\end{equation}
The normalization factor is 
\begin{equation}
    \mathcal{N}_{\vec{n},\mu} = \binom{N}{n_{-s},\ldots,n_s}\prod_{m\in\vec{m}_\mu} \big(n_m!\big)^2 ,
\end{equation}
which when using the multinomial coefficient,
\begin{equation}
    \binom{N}{n_1,n_2,\ldots,n_p}=\frac{N!}{n_1!\ n_2! \ldots n_p!\ \Big(N-\sum_{i=1}^p n_i\Big)!}\, ,
\end{equation}
simplifies to
\begin{equation}
     \mathcal{N}_{\vec{n},\mu} = N! \ n_{-s}! \ldots n_{\mu-1}!\ n_{\mu-1}! \ldots n_{s}!/ n_\mu! .
\end{equation}

It is particularly convenient to define GPES with respect to sea of maximally polarized state $|s\rangle$:
\begin{equation}
    |\vec{n},s \rangle \equiv |\vec{n} \rangle 
    = \mathcal{N}_{\vec{n}}^{-1/2}
    \prod_{m=-s}^{s-1} \left( \hat{S}^{s \rightarrow m} \right)^{n_m} |s \rangle ,
\end{equation}
with
\begin{equation}
    \mathcal{N}_{\vec{n}} = \binom{N}{n_{-s},\ldots,n_{s-1}}\prod_{m=-s}^{s-1} \big(n_m!\big)^2 .
\end{equation}
Throughout the following, we adopt the convention that the sea index $\mu$ is set to $s$ whenever it is not explicitly specified; thus, the maximally polarized sea $\mu=s$ serves as the default reference state.

The PES \eqref{eq:PES} obey orthogonality relation,
\begin{equation}
    \langle \vec{n},\mu| \vec{n}',\mu \rangle = \delta_{n_s,n_s'}\ldots\delta_{n_{-s},n_{-s}'}\equiv \delta_{\vec{n},\vec{n}'} .
\end{equation}
The populations satisfy the constraint $\sum_m n_m=N$, corresponding to the fixed total number of atoms.

The total number of GPES is:
\begin{equation}
    \#\{\text{GPES}\} =\overline{C}^N_{d}=\frac{(N+2s)!}{N!\,(2s)!} 
    \, .
\end{equation}
It is the $N$-element combination with repetitions of $d=2s+1$-element set: the number of ways in which one can choose magnetizations for $N$ atoms from $d=2s+1$-element set of possible magnetizations.

\subsection{$q$-wave population eigenstates ($q$-PES)}

We define the \emph{$q$-wave population eigenstate} ($q$-PES) as a PES characterized by a quasi-momentum $q$. Such states are obtained by acting on GPESs with the $q$-resolved S-arrow operator, which we will refer to as~\mbox{$q$-S-arrow} operator:
\begin{equation}
    \hat{S}^{m\rightarrow m'}_q = \sum_{j=1}^N e^{\mathrm{i}q\lambda j}\, \hat{S}^{m\rightarrow m'}_j.
\end{equation}
Here, $q$  denotes the quasi-momentum and $\lambda$ the lattice spacing between neighboring sites.
Under periodic boundary conditions (PBC), the allowed quasi-momenta are given by:
\begin{equation}
    q=2\pi\frac{k}{N}\lambda^{-1},\quad  k\in\{1,2,\ldots,N-1\}
\end{equation}
The $q$-PES $|q,m_q;\vec{n}\rangle$, is a population eigenstate in which the atoms transferred to the magnetic sub-level $m_q$ carry quasi-momentum $q$, while the remaining populations are described by the vector $\vec{n}'$. It is generated from a GPES by the action of the corresponding $q$-S-arrow operator in the following way:
\begin{equation}
    |q,m_q;\vec{n}'\rangle = (\mathcal{Q}_{n_{m_q}}^{n_{m'}})^{-1/2}\,\hat{S}^{{m'}\rightarrow m_q}_q\,|\vec{n}\rangle\label{eq:q-PES} .
\end{equation}
The $q$-S-arrow operator acts on a PES analogously to a collective S-arrow operator: it transfers one atom from the magnetic sub-level ${m'}$ to the target sub-level $m_q$, thereby increasing $n_{m_q}$ by one and decreasing $n_{m'}$ by one. Consequently, the population vector transforms as
\begin{equation}
    \vec{n}'=(n_s,\ldots,n_{m'}-1,\ldots,n_{m_q}\!+1,\ldots,n_{-s}) .
\end{equation}
The normalization factor associated with the action of the $q$-S-arrow operator is
\begin{equation}
    \mathcal{Q}_{n_{m_q}}^{n_{m'}}=\frac{n_{m'}(N-n_{m_q}\!-1)}{N-1}.
\end{equation}

The $q$-PES with a given $\vec{n}'$ can be also defined without involving relative GPES, by the direct generation from the sea: 
\begin{equation}
    |q,m_q;\vec{n}'\rangle = \mathcal{N}_{q, \vec{n}'\!,\mu}^{-1/2}
    \!\prod_{m\in\vec{m}_\mu}\!\! \left( \hat{S}^{\mu \rightarrow m} \right)^{n_m'\!-\delta_{m,m_q}} \hat{S}^{\mu\rightarrow m_q}_q\ |\mu\rangle ,\label{eq:q-PES_from_mu}
\end{equation}
where the normalization factor reads
\begin{equation}
    \mathcal{N}_{q, \vec{n}'\!,\mu} = 
    \frac{N-n_{m_q}'}{n_{m_q}'(N-1)}\ 
    \binom{N}{n_{-s}',\ldots,n_s'}\prod_{m\in\vec{m}_\mu} \big(n_m'!\big)^2.
\end{equation}
The $q$-PES defined by Eq.~\eqref{eq:q-PES_from_mu} is equivalent to the one introduced in Eq.~\eqref{eq:q-PES}. For a fixed choice of sea $\mu$, all $q$-PESs can be constructed except those with $m_q=\mu$, which require choosing a different reference sea.

There is one limitation for possible population configurations $\vec{n}'$ for $q$-PESsthey cannot correspond to a sea state, i.e., a configuration in which all atoms occupy a single magnetic sub-level, ($\vec{n}'\neq (0,\ldots,N,\ldots,0)$).In this case, the phase factors entering the $q$-S-arrow construction lead to complete cancellation of all contributions, and the resulting state vanishes.

An energy of the $q$-PES is dependent only on the quasi-momentum $q$:
\begin{equation}
    E_q = 2\,\mathrm{J}_\mathrm{SE}\Big(\cos q\lambda -1\Big).
\end{equation}
The $q$-PES are orthogonal in populations $\vec{n}$ and in quasi-momenta $q$,
\begin{equation}
    \langle q,m_q; \vec{n}| 
    q'\!, m_{q}; \vec{n}' \rangle = \delta_{q,q'}\,\delta_{\vec{n},\vec{n}'}.
\end{equation}

However, $q$-PESs corresponding to different magnetic sub-levels $m_q$ carrying the quasi-momentum $q$ are not, in general, orthogonal:
\begin{multline}
    \langle q,m_q; \vec{n}| 
    q, m_{q}'; \vec{n}\rangle =\\[3pt] = -\sqrt{\frac{n_{m_q}n_{m_q'}}{(N-n_{m_q})(N-n_{m_q'})}}(1-\delta_{m_q,m_q'})+\delta_{m_q,m_q'}.\label{eq:q-PES_non-ort}
\end{multline}
The non-orthogonality of $q$-PESs implies that they do not constitute an orthogonal basis within the subspace of states with fixed $q$ and $\vec{n}$. In fact, the set of $q$-PESs sharing the same $q$ and $\vec{n}$, but differing in the magnetic sub-level $m_q$, forms an overcomplete representation of the subspace that it spans:
\begin{equation}
    0=\sum\limits_{m=-s}^s \sqrt{n_m(N-n_m)}\ \Big|q,m_q{=}m,\vec{n}\Big\rangle\label{eq:qPES_overcompleat_1}
\end{equation}
Namely, every $q$-PES from this set can be expressed as a superposition of all other $q$-PES from the set:
\begin{multline}
    \Big|q,m_q{=}\tilde m,\vec{n}\Big\rangle =
    \frac{-1}{\sqrt{n_{\tilde m}(N-n_{\tilde m})}}\cdot\\
    \cdot\sum\limits_{m=-s}^s (1-\delta_{m,\tilde m})\sqrt{n_m(N-n_m)}\ \Big|q,m_q{=}m,\vec{n}\Big\rangle\label{eq:qPES_overcompleat_2}.
\end{multline}
This demonstrates that removing any one state from this set is sufficient to obtain a complete basis. In the following, we choose to exclude the state with $m_q=\mu$, which is the most convenient choice for our construction.

Once the completeness criterion for $q$-PES is established, one can count the number of linearly independent $q$-PES for given $q$:
\begin{multline}
    \#\{\text{$q$-PES}\}=\sum\limits_{k=2}^{d}\,\overline{C}^{N-k}_{k} \,C_{d}^{k}\,(k-1)=\\[3pt]
    =d!\, (N-1)!\sum\limits_{k=2}^{d}\frac{1}{(k-2)!\, k!\,(N-k)!\,(d-k)!},
\end{multline}
with $d=2s+1$.
For a given $\vec{n}$ the number of linearly independent possible $q$-PES is equal the number of levels with non-zero populations $k$ minus one for completeness (see the note under \eqref{eq:qPES_overcompleat_2}): $k=\sum_{m=-s}^s \sgn n_m$. Values of $k$ are integers from $2$ to $d$ (for $q$-PES at least two levels are occupied). Then, for given $k$ there is $C_{d}^{k}$ number of ways to choose $k$ levels that are occupied out of total $d$ number of levels. For a specific choice of occupied levels there are $\overline{C}^{N-k}_{k}$ number of ways of distributing $N$ atoms over $k$ levels provided that each of $k$ levels is occupied by at least one atom (where $\overline{C}$ stands for combination with repetitions).

\section{Spin Eigenstates of $\hat H_{\mathrm{SE}}$ \label{sec:SpinStates}}

All the PES states are the eigenstates of global $\hat J_z$ with magnetization $M=\sum_{m=-s}^s m\,n_m$:
\begin{align}
    \hat{J}_z |\vec{n} \rangle &= \left(\sum\limits_{m=-s}^s m\,n_m\ \right)|\vec{n}\rangle \\
    \hat{J}_z |q, m_q; \vec{n} \rangle &= \left( \sum\limits_{m=-s}^s m\,n_m\  \right)|q, m_q; \vec{n} \rangle.
\end{align}
In general, PESs are not eigenstates of the collective-spin operator $\hat{J}^2$. 
For GPESs, one finds:
\begin{equation}
    \hat{J}_{+} |\vec{n} \rangle = \sum_{m=-s}^{s-1}\! \alpha_{s,m}\sqrt{n_m(n_{m+1}+1)}\ |\vec{n}_{m}\rangle.
    \label{eq:J+onGPES}
\end{equation}
Here, $\vec{n}_m$ denotes the population vector obtained from $\vec{n}$ by transferring one atom from the magnetic sub-level $m$ to $m+1$,
\begin{equation}
    \vec{n}_{m} = \big(n_s,\ldots,n_{m+1}+1,n_{m}-1,\ldots,n_{-s}\big).
\end{equation}
For $q$-PESs, the action of the collective spin raising operator gives
\begin{align}
    \hat{J}_{+} \big|q, m_q; \vec{n} \big\rangle 
    &=\!\sum_{m=-s}^{s-1}\! F_{m}^{m_q,\vec{n}}\, \big|q, m_q; \vec{n}_m \big\rangle\ +\nonumber\\[3pt] &+\, f^{m_q,\vec{n}}\, \big|q, m_q+1; \vec{n}_{m_q} \big\rangle\label{eq:J+onqPES}
\end{align}
where the corresponding coefficients are:
\begin{align}
    F_{m}^{m_q,\vec{n}} &= \alpha_{s,m}\sqrt{n_m(n_{m+1}+1)}\ \,\frac{n_m-\delta_{m_q,m}}{n_m}\cdot  \nonumber\\[3pt]
     &\phantom{=\ }\cdot\sqrt\frac{n_{m_q}(N-n_{m_q}-\delta_{m_q-1,m}+\delta_{m_q,m})}{(N-n_{m_q})(n_{m_q}+\delta_{m_q-1,m}-\delta_{m_q,m})}, \\[7pt]
    f^{m_q,\vec{n}} &= \alpha_{s,m_q}\sqrt\frac{N-n_{m_q+1}-1}{N-n_{m_q}}. 
\end{align}

Since $\hat J^2 = \hat J_z^2 + \hat J_z + \hat J_- \hat J_+$, Eqs.~\eqref{eq:J+onGPES} and \eqref{eq:J+onqPES} immediately imply that PESs are, in general, not eigenstates of the collective-spin operator  $\hat J^2$. Indeed, the action of  $\hat J_+$ on a single PES produces a superposition of multiple PESs rather than a single state. This result is consistent with the more general observation discussed above: the collective S-arrow operators, including the population operators  
$\hat n$, do not commute with the collective-spin operators, preventing PESs from being simultaneous eigenstates of both sets of observables.

\subsection{General construction of total spin eigenstates \label{ssec:SpinStates_GeneralConstruction}}

We denote simultaneous eigenstates of
$\hat H_{\rm SE}$, total spin $\hat J^2 $ and $\hat J_z$ by $|E,\Lambda,M\rangle$, where $\Lambda$ and $M$ are the collective-spin and collective-magnetization quantum numbers, respectively. These states satisfy
\begin{equation}
    \hat J^2\,|\Lambda,M\rangle=\Lambda(\Lambda+1)\,|\Lambda,M\rangle .
\end{equation}

For a given energy manifold ($E=0$ or $E_q$) the spin state can be expressed in therms of PES states of energy $E$ and such population sets $\vec{n}$ which gives magnetization $M$: 
\begin{equation}
|E,\Lambda,M\rangle=\sum\limits_{\vec{n}\in\{\vec n\}(M)}\!\sum\limits_{f} a_{\vec{n},f}\,|E,\vec{n},f\rangle\label{eq:|L,M>inPES},
\end{equation}
where $f$ is a superindex taking into account any further possible PES degeneration (for example for $q$-PES it could be $m_q$) and  $\{\vec n\}(M)$ is a set of all collective populations of magnetic sub-levels $\vec{n}$ that give collective magnetization equal to $M$, namely $M=\sum_{m=-s}^s m\,n_m$. 

Starting from any state $|E,\Lambda,M\rangle$, the remaining members of the same collective-spin multiplet can be generated through successive applications of the ladder operators $\hat J_\pm$. Consequently, it is sufficient to determine a single state for each multiplet. For our purposes, it is particularly convenient to consider the highest-weight state $|E,\Lambda,\Lambda\rangle$, which has the maximal magnetization $M=\Lambda$. Such a state satisfies
\begin{equation}
    \hat J_+|E,\Lambda,\Lambda\rangle=0\label{eq:J+onMmax=zero}
\end{equation}
and therefore provides a natural starting point for constructing the entire multiplet.

We refer to such a highest-weight state as the parent state of the multiplet. Expanding $|E,\Lambda,\Lambda\rangle$ in the PES basis according to Eq.~\eqref{eq:|L,M>inPES}, applying $\hat J_+$ via Eq.~\eqref{eq:J+onGPES}, and collecting terms corresponding to the same $\vec{n}$ and $f$ we obtain
\begin{equation}
    \hat J_+\,|E,\Lambda,\Lambda\rangle=
 \sum\limits_{\vec{n}}\sum\limits_f C_{\vec{n},f}\,|E,\vec{n},f\rangle = 0.
\end{equation}
Because the chosen PES basis is linearly independent (a nontrivial requirement in the case of $q$-PESs), the highest-weight condition $\hat J_+\,|E,\Lambda,\Lambda\rangle= 0$
implies
\begin{equation}
    C_{\vec{n},f}=0,\quad
    \text{where}\quad 
    \vec{n}\in\{\vec n\}(M{=}\Lambda{+}1)\, .
    \label{eq:C_n}
\end{equation}
The coefficients $C_{\vec{n},f}$ are linear combinations of the expansion coefficients $a_{\vec{n},f}$ appearing in Eq.~\eqref{eq:|L,M>inPES}. Consequently, the above conditions define a system of linear equations for $a_{\vec{n},f}$, whose solutions determine the parent state $|E,\Lambda,\Lambda\rangle$ in the PES representation. Different choices of independent solutions generally lead to different families of collective-spin eigenstates.

To address this point, let us examine the structure of the parent state $|E,\Lambda,\Lambda\rangle$ in its PES decomposition as in Eq.~\eqref{eq:|L,M>inPES}. The population vectors $\vec{n}$ entering the expansion are constrained only by two conditions: conservation of magnetization, $\sum_m m\, n_m = \Lambda$, and conservation of particle number, $\sum_m n_m =N$. 
A natural question then arises: does the construction of the parent state $|E,\Lambda,\Lambda\rangle$ require PESs corresponding to all population configurations $\vec n$ satisfying these constraints, or only a subset thereof?

The answer follows from the structure of the raising operator $\hat{J}_+$, crucial for calculation. Since $\hat{J}_+$ is composed of operators $S^{m\to m+1}$), its action can only transfer population from lower to higher magnetic sub-levels. Consequently, if a PES $|E,\vec{n}_\zeta,f\rangle$ occupies only the $\zeta$ highest magnetic sub-levels,
\begin{equation}
    \vec{n}_\zeta = (\underbrace{n_s,\ldots,n_{s-\zeta+1}}_\zeta ,0,\ldots,0),
    \label{eq:n_zeta}
\end{equation}
the action of $\hat{J}_+$ cannot populate any of the remaining sub-levels. Therefore, the subspace spanned by such states is closed under $\hat{J}_+$ . 
follows that, irrespective of the total number of magnetic sub-levels $d$, parent states may be constructed using only the $\zeta$ highest sub-levels, with $\zeta\leq d$. This naturally partitions the parent states into families labeled by $\zeta=1,2,\ldots,d$ which we refer to as the parent-component number. The corresponding parent states take the form
\begin{equation}
    |E,\Lambda,\Lambda\rangle_\zeta
    =\sum\limits_{\vec{n}_\zeta\in\{\vec n\}(\Lambda)}\sum\limits_f a_{\vec{n}_\zeta,f}\,|E,\vec{n}_\zeta,f\rangle
    \label{eq:LLzeta}
\end{equation}
Each value of $\zeta$ defines a distinct family of collective-spin eigenstates characterized by the energy $E$, spin length $\Lambda$, and parent-component number $\zeta$. 

In principle, up to $d$ distinct families of states can be constructed, corresponding to the possible values of $\zeta=1,\ldots,d$. However, the existence of a family for a given $\zeta$ is not guaranteed. Moreover, for a fixed $\zeta$, only specific values of the collective spin $\Lambda$ may be allowed. For example, as shown below, the $\zeta=1$ family contains only the Dicke states with $\Lambda=sN$.

The solution of the constraint equations \eqref{eq:C_n} is determined primarily by the value of $\zeta$ and is largely independent of the single-atom spin $s$. Therefore, the construction of collective-spin eigenstate families will be performed separately for each parent-component number $\zeta$ and for each PES energy sector considered in this work (GPESs or $q$-PESs).

It is important to note that families corresponding to different values of $\zeta$ are orthogonal. This follows from the orthogonality of their parent states: states with different $\zeta$ are constructed from PESs belonging to distinct, non-overlapping population sectors, and PESs with different population vectors are orthogonal.

\subsection{Total spin eigenstates for $E=0$}

Let us first consider the construction of ground spin eigenstates, i.e., states composed of GPESs. Since all such states have zero energy, $E=0$, we omit the energy index in the notation, as for GPESs. Moreover, for a given population vector $\vec n$ there is a unique GPES, so no additional index $f$ is required. Therefore, the general parent state in the $E=0$ sector takes the form:
\begin{equation}
    |\Lambda,\Lambda\rangle_\zeta
    =\sum\limits_{\vec{n}_\zeta\in\{\vec n\}(\Lambda)} a_{\vec{n}_\zeta}\,|\vec{n}_\zeta\rangle
    \label{eq:LL_GPES}
\end{equation}

\paragraph{$\zeta=1$ case:}

The $\zeta=1$ is a trivial case when the parent state is just a single GPES: 
\begin{equation}
    |\Lambda,\Lambda\rangle_{\zeta=1}
    =|\vec{n}_{\zeta=1}\rangle,\,\, {\rm with}
    \,\, \vec{n}_{\zeta=1}=(N,0,\cdots,0),
    \label{eq:LLzeta1}
\end{equation}
Since there is no degree of freedom left for $\zeta=1$ there is only one possible value of $\Lambda$ which is the maximal spin length of the system  $\Lambda = s N$.

The family of states generated from $|\Lambda,\Lambda\rangle_{\zeta=1}=|\text{max}\rangle$ parent state is called Dicke manifold:
\begin{equation}
    |\Lambda,M\rangle_{\zeta=1}=
    \sqrt{\frac{(\Lambda+M)!}{(\Lambda-M)!(2\Lambda )!}} \hat{J}_-^{\Lambda-M} 
    |\Lambda,\Lambda\rangle_{\zeta=1}.
\end{equation}
The Dicke family exists for any elementary spin $s$.

It is also important to notice that there are only two specific GPES that are spin states alone. These are two sea states $|s\rangle=|\vec{n}{=}(N,0,\ldots,0)\rangle$ and $|{-}s\rangle=|\vec{n}{=}(0,\ldots,0,N)\rangle$, that are respectively the maximally ($M=sN$) and the minimally ($M={-}sN$) polarized states of the system, hence in the context of spin states they are denoted as $|\text{max}\rangle$ and $|\text{min}\rangle$ respectively. The later is the parent for Dicke family of states $|\Lambda,\Lambda\rangle_{\zeta=1} =|s\rangle=|\text{max}\rangle$
the former could also play a role of parent state of Dicke family if we were defining it from the minimal, not the maximal, magnetization $|\Lambda,{-}\Lambda\rangle_{\zeta=1} =|{-}s\rangle=|\text{min}\rangle$

\paragraph{$\zeta=2$ case:}

It turns out that for $\zeta=2$ no $E=0$ energy spin state can be constructed for any $\Lambda$.

This is because for $\zeta=2$ there is always only one GPES with the required magnetization $\Lambda$. Namely:
\begin{equation}
    |\Lambda{=}sN{-}k,\Lambda\rangle_{\zeta=2}=
    a_{\vec{n}_{\zeta=2}}\Big|\vec{n}_{\zeta=2}{=}\big(N{-}k, k,0,\ldots,0\big)\Big\rangle
\end{equation}
That only GPES is not a spin state on its own (despite $k=0$ case, that is just the Dicke parent state, so $\zeta=1$ family actually), what can be easily checked. This means, that the only way to satisfy $\hat{J}_+|\Lambda,\Lambda\rangle_{\zeta=2}=0$ is setting $a_{\vec{n}_{\zeta=2}}=0$, what means, that $|\Lambda,\Lambda\rangle_{\zeta=2}$ does not exist, and hence there are no families of spin states for  $\zeta=2$ parent population.

\paragraph{$\zeta=3$ case:}

First non-trivial parent component number that allows for the constriction of some families is $\zeta=3$, when the three top magnetic sub-levels are populated, 
$\vec{n}_{\zeta=3}=(n_s, n_{s-1}, n_{s-2}, 0, \cdots, 0)$. 

Since every atom populating $n_{s-2}$ lowers global magnetization by $2$ with respect to the maximal $sN$, the consideration of $|\Lambda,\Lambda\rangle_{\zeta=3}$ states is held differently for $\Lambda$ lower than $sN$ by even and odd number:
\begin{align}
    |\Lambda&\!=\!sN\!-\!2k,\Lambda\rangle_{\zeta=3} =\nonumber\\[-3pt]
    &= \sum_{p=0}^k a_{\vec n(p)}\,\big|\vec n(p)\!=\!(n_{s},2(k\!-\!p),p,0,\ldots)\big\rangle\label{eq:zeta3_even}\\[10pt]
    |\Lambda&\!=\!sN\!-\!2k\!+\!1,\Lambda\rangle_{\zeta=3} =\nonumber\\[-3pt]
    &= \sum_{p=0}^{k-1} a_{\vec n(p)}\,\big|\vec n(p)\!=\!(n_{s},2(k\!-\!p)\!-\!1,p,0,\ldots)\big\rangle\label{eq:zeta3_odd}
\end{align}
Where $k=1,2,\ldots,\lfloor\frac{N}{2}\rfloor$ and $n_s$ is determined from total particle number being fixed $n_s=N-n_{s-1}-n_{s-2}$: in \eqref{eq:zeta3_even} it is $n_s=N-2k+p$ and in \eqref{eq:zeta3_odd} it is $n_s=N-2k+p+1$.

Let us focus on the case with $\Lambda=sN-2k$, since later it will turn out that it is the only physically relevant case.

Acting on the \eqref{eq:zeta3_even} with $J_+$ one obtains:
\begin{align}
    \hat J_+&\,|\Lambda\!=\!sN\!-\!2k,\Lambda\rangle_{\zeta=3} =\nonumber\\
    & \sum_{p=0}^{k-1} a_{\vec n(p)}\,\alpha_{s,s-1}\sqrt{2(k\!-\!p)(N\!-\!2k\!+\! p\!+\!1)}\,\cdot\nonumber\\[-5pt]
    &\hspace{40pt}\cdot\big|\vec n\!=\!(n_s\!+\!1,2(k\!-\!p)\!-\!1,p,0,\ldots)\big\rangle\nonumber\\
    +& \sum_{p=1}^k a_{\vec n(p)}\,\alpha_{s,s-2}\sqrt{p\big(2(k\!-\! p)\!+\!1\big)}\,\cdot\nonumber\\[-5pt]
    &\hspace{40pt}\cdot\big|\vec n\!=\!(n_s,2(k\!-\!p)\!+\!1,p\!-\!1,0,\ldots)\big\rangle
\end{align}
where $\alpha$ is defined in \eqref{eq:alpha}. Gathering together terms with the same GPES we obtain:
\begin{multline}
     \hat J_+\,|\Lambda\!=\!sN\!-\!2k,\Lambda\rangle_{\zeta=3} =\\= \sum_{p=0}^{k-1} C_{p}\, \big|\vec n\!=\!(n_s,2(k\!-\!p)\!-\!1,p,0,\ldots)\big\rangle
\end{multline}
with:
\begin{multline}
    C_p = a_{\vec n(p)}\,\alpha_{s,s-1}\sqrt{2(k\!-\!p)(N\!-\!2k\!+\! p\!+\!1)}\ +\\+\,a_{\vec n(p+1)}\,\alpha_{s,s-2}\sqrt{(p+1)\big(2(k\!-\! p)\!-\!1\big)}
\end{multline}
Since $|\Lambda,\Lambda\rangle$ has to be annihilated by $\hat J_+$, all $C_p$ have to be equal to zero, what gives a set of equations for $a_{\vec{n}}$ coefficients in the form of recursion relation:
\begin{equation}
    a_{\vec n(p)}=-\frac{\alpha_{s,s-2}}{\alpha_{s,s-1}}\sqrt\frac{(p+1)\big(2(k\!-\! p)\!-\!1\big)}{2(k\!-\!p)(N\!-\!2k\!+\! p\!+\!1)}\ a_{\vec n(p+1)}
\end{equation}
and overall magnitude of $a_{\vec{n}}$ are set by normalization of $|\Lambda,\Lambda\rangle$.

It remains to check if case described by eq.\ \eqref{eq:zeta3_odd} gives some solution. The procedure is analogues to \eqref{eq:zeta3_even} case. First we act on \eqref{eq:zeta3_odd} with $\hat J_+$ and demand that it gives zero:
\begin{align}
    \hat J_+&\,|\Lambda\!=\!sN\!-\!2k\!+\!1,\Lambda\rangle_{\zeta=3} =\nonumber\\
    & \sum_{p=0}^{k-1} a_{\vec n(p)}\,\alpha_{s,s-1}\sqrt{[2(k\!-\!p)\!-\!1](N\!-\!2k\!+\! p\!+\!2)}\,\cdot\nonumber\\[-5pt]
    &\hspace{40pt}\cdot\big|\vec n\!=\!(n_s\!+\!1,2(k\!-\!p)\!-\!2,p,0,\ldots)\big\rangle\nonumber\\
    +& \sum_{p=1}^{k-1} a_{\vec n(p)}\,\alpha_{s,s-2}\sqrt{p\big(2(k\!-\! p)\big)}\,\cdot\nonumber\\[-5pt]
    &\hspace{40pt}\cdot\big|\vec n\!=\!(n_s,2(k\!-\!p),p\!-\!1,0,\ldots)\big\rangle
\end{align}
After gathering together terms with the same GPES:
\begin{multline}
     \hat J_+\,|\Lambda\!=\!sN\!-\!2k\!+\!1,\Lambda\rangle_{\zeta=3} =\\= \sum_{p=0}^{k-1} C_{p}\, \big|\vec n\!=\!(n_s\!+\!1,2(k\!-\!p)\!-\!2,p,0,\ldots)\big\rangle
\end{multline}
where:
\begin{align}
    C_p &= a_{\vec n(p)}\,\alpha_{s,s-1}\sqrt{[2(k\!-\!p)\!-\!1](N\!-\!2k\!+\! p\!+\!2)}\ +\nonumber\\
    &\hspace{25pt}+\, a_{\vec n(p+1)}\,\alpha_{s,s-2}\sqrt{(p+1)\big(2(k\!-\! p)\!-\!2\big)}\label{eq:zeta3_odd_C_p}\\[3pt]
   &\hspace{70pt} \text{for}\qquad p=0,1,\ldots,k\!-\!2\nonumber\\[5pt]
    C_{k-1}&=a_{\vec n(k-1)}\,\alpha_{s,s-2}\sqrt{N-k+1}\label{eq:zeta3_odd_C_k-1}
\end{align}
The $\hat{J}_+|\Lambda\!=\!sN\!-\!2k\!+\!1,\Lambda\rangle_{\zeta=3} =0$ condition is met only when all $C_p$ coefficients are zero. In particular, setting $C_{k-1}=0$ gives from \eqref{eq:zeta3_odd_C_k-1}, that $a_{\vec n(k-1)}$ hast to be identically zero.  If  $a_{\vec n(k-1)}=0$ then from \eqref{eq:zeta3_odd_C_p} it follows that all other $a_{\vec n(p)}$ coefficients have to be zero, so the $|\Lambda\!=\!sN\!-\!2k\!+\!1,\Lambda\rangle_{\zeta=3} $ state does not exist.

Concluding the $\zeta=3$ class states for $E=0$, it turns out that families of states exist only for $\Lambda=sN-2k$ (with $k=1,2,\ldots,\lfloor\frac{N}{2}\rfloor$) while for all $\Lambda=sN-2k+1$ the collective spin states do not exist in this $\zeta$ class.

\paragraph{$\zeta=4$ case:}

Authors also considered the case $\zeta=4$. Although the construction becomes significantly more computationally demanding, it remains feasible and yields collective-spin families with $\Lambda=sN-3k$ and $\Lambda=sN-3k-2$, where $k$ is a non-negative integer. No families are found for $\Lambda=sN-3k-1$. The resulting expressions for the coefficients $a_{\vec n}$ are, however, cumbersome and do not provide additional physical insight. Therefore, we do not present these explicit forms and restrict the discussion to the observation that the construction of such families is possible for the corresponding allowed values of $\Lambda$.

\subsection{Generalized Spin Wave states}

Construction of spin states in the $E_q$ manifold is also done according to section \ref{ssec:SpinStates_GeneralConstruction}, heaving in mind that compared to the simple case of $E=0$ manifold this time $m_q$ plays the role of superindex $f$ and also that one value of $m_q$ has to be excluded to provide linear independence of $q$-PES involved (in our case we choose $m_q\neq s$):
\begin{equation}
    |q,\Lambda,\Lambda\rangle_\zeta
    =\sum\limits_{\vec{n}_\zeta\in\{\vec n\}(\Lambda)}\ \sum\limits_{m_q=-s}^{s-1} a_{\vec{n}_\zeta}^{m_q}\,|q,m_q,\vec{n}_\zeta\rangle
    \label{eq:LLzeta_qPES}
\end{equation}

\paragraph{$\zeta=1$ case} does not exist, since there are not any $q$-PES states with only one magnetic sub-level populated (no sea $q$-PES).

\paragraph{$\zeta=2$ case:} Analogously to the $E=0$ manifold, for the required magnetization $\Lambda$ there is only one $\vec{n}_{\zeta=2}$, and since only two magnetic sub-levels are  occupied, there is only one possible value of $m_q$, namely $m_q=s-1$:
\begin{multline}
    |q,\Lambda{=}sN{-}k,\Lambda\rangle_{\zeta=2}\overset{?}{=}\\[3pt]
    \overset{?}{=}\Big|q,m_q{=}s{-}1,\vec{n}_{\zeta=2}{=}\big(N{-}k, k,0,\ldots,0\big)\Big\rangle
\end{multline}
However, this time for $k=1$ ($\Lambda= sN{-}1$) case, such a $q$-PES is a spin state (what can be easily checked by acting with $\hat J_+$ on it):
\begin{multline}
    |q,\Lambda{=}sN{-}1,\Lambda\rangle_{\zeta=2}=\\[3pt]
    =\Big|q,m_q{=}s{-}1,\vec{n}{=}\big(N{-}1, 1,0,\ldots,0\big)\Big\rangle
\end{multline}

The $|q,\Lambda{=}sN{-}1,\Lambda\rangle_{\zeta=2}$ state is a parent state for a family of states called \emph{Spin-Wave States} (SWS), which are known as eigenstates of the isotropic XXX Heisenberg model.

There are only two linearly independent $q$-PESs that are collective-spin eigenstates by themselves: $|q,\Lambda{=}sN{-}1,\Lambda\rangle=|q,m_q{=}s{-}1;\vec{n}{=}(N{-}1,1,0,\ldots,0)\rangle$ and $|q,\Lambda{=}sN{-}1,{-}\Lambda\rangle=|q,m_q{=}{-}s;\vec{n}{=}(0,\ldots,0,1,N{-}1)\rangle$. Both states belong to the same SWS family. The first one is the parent state introduced above, constructed from the maximal-magnetization sector. The second can be interpreted as an alternative parent state if the family is defined starting from the minimal rather than maximal magnetization state.

For spin-$1/2$ particles, the SWS family contains all possible spin eigenstates in the $E_q$ sector, as $\zeta=2=d$.

\paragraph{$\zeta=3$ case:}
Two possible classes of $\Lambda$ require separate treatment: one is $\Lambda=sN-2k$, other is $\Lambda=sN-2k-1$, where $k=1,2,\ldots,\lfloor\frac{N}{2}\rfloor$. Let's attack the firs class. The general form of state in compleat description in $q$-PES is:
\begin{align}
    \Big|q,&\Lambda{=}sN{-}2k,M{=}\Lambda\Big\rangle_{\zeta=3}=\nonumber\\
    &= \sum_{p=0}^{k-1} b_{\vec{n}_{\mathrm{even}}(p)}^{m_q=s-1}\Big|q,m_q{=}s{-}1,\vec{n}_{\mathrm{even}}(p)\Big\rangle+\nonumber\\
    &\phantom{=}+\sum_{p=1}^{k} b_{\vec{n}_{\mathrm{even}}(p)}^{m_q=s-2}\Big|q,m_q{=}s{-}2,\vec{n}_{\mathrm{even}}(p)\Big\rangle\label{eq:|q,L,M>_even}
\end{align}
where $\vec{n}_{\mathrm{even}}(p)=\big(N-2k+p,2(k-p),p,0,\ldots\big)$. The states $\big|q,m_q{=}s,\vec{n}\big\rangle$ were excluded from decomposition for a manner of completeness. 

Acting with $\hat J_+$ on the ansatz \eqref{eq:|q,L,M>_even} we obtain the state which is required to be zero. From the condition of zeroing its coefficients, we obtain a set of $2k-1$ equations, which let us recursively construct the $b$ coefficients (there are $2k$ coefficients $b$ so they are specified by the set of $2k-1$ equation up to the global normalization factor):
\begin{widetext}
\begin{align}
    b_{\vec{n}_{\mathrm{even}}(k-1)}^{m_q=s-1}&=\frac{- f^{s-2,\vec{n}_{\mathrm{even}}(k)}\  b_{\vec{n}_{\mathrm{even}}(k)}^{m_q=s-2}}{ F_{s-1}^{s-1,\vec{n}_{\mathrm{even}}(k-1)}- f^{s-1,\vec{n}_{\mathrm{even}}(k-1)}\sqrt{\tfrac{N-1}{(N-k)k}}}\label{eq:|q,L,M>_b_max}\\[10pt]
    b_{\vec{n}_{\mathrm{even}}(p)}^{m_q=s-2}&=\frac{-\Big(F_{s-2}^{s-2,\vec{n}_{\mathrm{even}}(p+1)}\ b_{\vec{n}_{\mathrm{even}}(p+1)}^{m_q=s-2} - f^{s-1,\vec{n}_{\mathrm{even}}(p)}\sqrt{\tfrac{p(N-p)}{(N-2k+p+1)(2k-p-1)}}\ b_{\vec{n}_{\mathrm{even}}(p)}^{m_q=s-1}\Big)}{F_{s-1}^{s-2,\vec{n}_{\mathrm{even}}(p)}}\label{eq:|q,L,M>_b_s-2}\\[10pt]
    b_{\vec{n}_{\mathrm{even}}(p)}^{m_q=s-1}&=\frac{-\Big(F_{s-2}^{s-1,\vec{n}_{\mathrm{even}}(p+1)}\ b_{\vec{n}_{\mathrm{even}}(p+1)}^{m_q=s-1} + f^{s-2,\vec{n}_{\mathrm{even}}(p+1)}\ b_{\vec{n}_{\mathrm{even}}(p+1)}^{m_q=s-2}\Big)}{F_{s-1}^{s-1,\vec{n}_{\mathrm{even}}(p)}-f^{s-1,\vec{n}_{\mathrm{even}}(p)}\sqrt{\tfrac{(2(k-p)-1)(N-2(k-p)+1)}{(N-2k+p+1)(2k-p-1)}}}\label{eq:|q,L,M>_b_s-1}
\end{align}
\end{widetext}
where in equation \eqref{eq:|q,L,M>_b_s-2} parameter $p=k-1,\ldots,1$, while in  \eqref{eq:|q,L,M>_b_s-1} parameter $p=k-2,\ldots,0$. First equation \eqref{eq:|q,L,M>_b_max} gives us relation between two $b$ coefficients with maximal values of $d$: $b_{\vec{n}(k-1)}^{m_q=s-1}$ and $b_{\vec{n}(k)}^{m_q=s-2}$, then using sequently equations \eqref{eq:|q,L,M>_b_s-2}, \eqref{eq:|q,L,M>_b_s-1} and the again  \eqref{eq:|q,L,M>_b_s-2} and \eqref{eq:|q,L,M>_b_s-1} \textit{etc.} we reproduce relative values of all $b$ coefficients in \eqref{eq:|q,L,M>_even}.

Procedure is analogs for $\Lambda=sN-2k-1$ where $k=1,2,\ldots,\lfloor\frac{N}{2}\rfloor$:
\begin{align}
    \Big|q,&\Lambda{=}sN{-}2k{-}1,M{=}\Lambda\Big\rangle_{\zeta=3}=\nonumber\\
    &= \sum_{p=0}^{k} c_{\vec{n}_{\mathrm{odd}}(p)}^{m_q=s-1}\Big|q,m_q{=}s{-}1,\vec{n}_{\mathrm{odd}}(p)\Big\rangle+\nonumber\\
    &\phantom{=}+\sum_{p=1}^{k} c_{\vec{n}_{\mathrm{odd}}(p)}^{m_q=s-2}\Big|q,m_q{=}s{-}2,\vec{n}_{\mathrm{odd}}(p)\Big\rangle\label{eq:|q,L,M>_odd}
\end{align}
where $\vec{n}_{\mathrm{odd}}(p)=\big(N-2k+p-1,2(k-p)+1,p,0,\ldots\big)$ (case $k=0$ would lead to a $\zeta=2$ states -- the SWS that were discussed before).

We find the set of equations for $c_{\vec{n}_{\mathrm{odd}}}^{m_q}$ coefficients:
\begin{widetext}
    \begin{align}
        c_{\vec{n}_{\mathrm{odd}}(k)}^{m_q=s-2}&=\frac{-f^{s-1,\vec{n}_{\mathrm{odd}}(k)}\ c_{\vec{n}_{\mathrm{odd}}(k)}^{m_q=s-1}}{F_{s-1}^{s-2,\vec{n}_{\mathrm{odd}}(k)}} \\[5pt]
        c_{\vec{n}_{\mathrm{odd}}(d)}^{m_q=s-1} &= \frac{-\Big( F_{s-2}^{s-1,\vec{n}_{\mathrm{odd}}(d+1)}\ c_{\vec{n}_{\mathrm{odd}}(d+1)}^{m_q=s-1}+f^{s-2,\vec{n}_{\mathrm{odd}}(d+1)}\ c_{\vec{n}_{\mathrm{odd}}(d+1)}^{m_q=s-2}\Big)}{ F_{s-1}^{s-1,\vec{n}_{\mathrm{odd}}(d)}-f^{s-1,\vec{n}_{\mathrm{odd}}(d)}\sqrt{\tfrac{2(k-d)(N-2(k-d))}{(N-2k+d)(2k-d)}}} \\[5pt]
        c_{\vec{n}_{\mathrm{odd}}(d)}^{m_q=s-2} &= \frac{-\Big( F_{s-2}^{s-2,\vec{n}_{\mathrm{odd}}(d+1)}\ c_{\vec{n}_{\mathrm{odd}}(d+1)}^{m_q=s-2} - f^{s-1,\vec{n}_{\mathrm{odd}}(d)}\sqrt{\tfrac{d(N-d)}{(N-2k+d)(2k-d)}}\ c_{\vec{n}_{\mathrm{odd}}(d)}^{m_q=s-1} \Big)}{F_{s-1}^{s-2,\vec{n}_{\mathrm{odd}}(d)}}
    \end{align}
\end{widetext}

We see that in $\zeta=3$ class there exist a family of $E_q$ spin states for every collective spin length from the set: $\Lambda=sN,sN-1,sN-2,\ldots,sN-N$ in contrary to the $E=0$ case that in this class of $\zeta=3$ can take only $\Lambda$s diminished from maximal by even number: $\Lambda=sN, sN-2, sN-4,\ldots,sN-2 \lfloor \frac{N}{2}\rfloor$.

\subsection{Spin eigenstate families: summary}

The existence of spin eigenstates of the SE Hamiltonian has been demonstrated explicitly for several representative cases. A general construction procedure is provided in Sec.~\ref{ssec:SpinStates_GeneralConstruction}; however, since each pair of parameters $\zeta$ and $\Lambda$ requires a separate analysis, no universal closed-form solution can be obtained.

Moreover, even in the cases where the construction is possible, the resulting spin eigenstates expressed in the PES basis (and therefore in the Fock basis) have a highly involved structure, which limits their practical use in calculations. Their construction should therefore be viewed primarily as a proof of principle, confirming that the commutation relation between $\hat{H}_{\mathrm{SE}}$ and $\hat{\vec J}$ allows the existence of common eigenstates of $\hat{H}_{\mathrm{SE}}$, $\hat J^2$, and $\hat J_z$.

There is, however, an important lesson from finding spin eigenstates of $\hat{H}_{\mathrm{SE}}$. Namely, the knowledge of possible values of collective spin $\Lambda$ in specific energy manifolds. 

For the $E=0$ manifold and elementary spin $s=1/2$, the only collective-spin eigenstates belong to the Dicke manifold with the maximal spin value $\Lambda=sN$. In contrast, already for $s=1$, multiple families of spin eigenstates appear, with $\Lambda$ ranging from the maximal value $\Lambda=sN$ through values reduced by even integers, $\Lambda=sN-2k$, down to the lowest allowed values. 

A similar behavior occurs in the $E_q$ manifold. For $s=1/2$, the SWS family exhausts all spin eigenstates with energy $E_q$, corresponding to $\Lambda=sN-1$. For $s\geq1$, however, the same energy manifold contains spin eigenstates with a broad range of collective-spin values, extending from $\Lambda=sN-1$ to substantially lowest values $\Lambda=0$.

These differences have profound consequences for the spin dynamics generated by the SE Hamiltonian. In the $s=1/2$ case, fixing the energy manifold also fixes the collective spin: $E=0$ corresponds uniquely to the fully polarized state with $\Lambda=sN$, while $E_q$ corresponds to $\Lambda=sN-1$. For $s\geq1$, the same energy manifolds contain states with many different collective-spin lengths, fundamentally modifying spin dynamics.

This enhanced spin degeneracy for $s\geq1$ originates from the symmetry of the SE Hamiltonian in the space of atomic magnetic sub-levels. The Hamiltonian possesses an $\mathrm{SU}(d)$ symmetry, reflecting its invariance under arbitrary transformations mixing the $d=2s+1$ magnetic sub-levels. In contrast, the isotropic Heisenberg Hamiltonian possesses only global $\mathrm{SU}(2)$ symmetry, corresponding to invariance under collective rotations of the spin degrees of freedom.

For $s=1/2$, the two symmetries coincide because $d=2$, and the spin-exchange Hamiltonian becomes equivalent to the Heisenberg model. This equivalence is broken for all $s\geq1$, leading to qualitatively different many-body dynamics and spin properties.

This multi-$\Lambda$ structure within each energy manifold has direct dynamical consequences: any perturbation coupling different energy sectors will generically drive population transfer between states of different collective spin length, even when the initial state has a well-defined $\Lambda$. This is the microscopic origin of the breakdown of fixed-spin-manifold descriptions for $s \geq 1$ as discussed in the next Section.

\section{Coupling to light}

We concluded that the practical usage of spin eigenstates of SE Hamiltonian is rather limited. We would like to confirm the correctness of the PES framework by calculation of dynamics employing $\hat{H}_{\mathrm{SE}}$, and hopefully presenting the influence of atomic spin length $s$ on the collective spin dynamics.

In particular, we are interested in the spin squeezing dynamics in the system. 
Spin-squeezed states are many-body states where fluctuations of a collective-spin component are reduced below the quantum noise level, with the reduction occurring at the expense of increased fluctuations in the conjugate component. 
The dynamics starts with a coherent spin state which is obtained by applying a collective rotation to the maximally polarized state $|\mathrm{max}\rangle$,
\begin{equation}
    |\vartheta,\varphi\rangle = e^{-\mathrm{i} \hat{J}_z \varphi} e^{-\mathrm{i} \hat{J}_y \vartheta}|\mathrm{max}\rangle.
\end{equation}
Squeezing level is measured by the spin squeezing parameter~\cite{Wineland1992, Wineland1994}:
\begin{equation}
    \xi^2=2sN\,\Delta J_{\perp,\mathrm{min}}^2/\langle\hat J\rangle^2.
\end{equation}
It quantifies the minimal transverse collective-spin uncertainty, $\Delta J_{\perp,\mathrm{min}}$, relative to the mean collective spin length. Spin squeezing occurs when this parameter falls below unity, indicating phase estimation beyond the standard quantum limit.
The focus of the present work is the construction and validation of the PES framework; a detailed analysis of the squeezing dynamics and its metrological implications for varying system parameters is presented in the companion paper~\cite{short}.

It was shown in Ref.~\cite{Tana2022} that for spin-$1/2$ fermions in an optical lattice weakly coupled to light, via the spin-orbit interaction $\hat{V}{\mathrm{soc}}$, can generate scalable spin squeezing by effectively realizing the One-Axis Twisting (OAT) model. The latter is described by the Hamiltonian $\hat{H}_{\mathrm{OAT}}=\chi \hat{J}_\sigma^2$, which transforms an initial spin coherent state into a spin-squeezed state. In Ref.~\cite{Domantas2024}, we showed how various couplings between the nuclear spin of alkaline-earth atoms and light can be engineered, including the spin-orbit coupling.

Here, we consider spin-orbit coupling in the form \cite{Tana2022}:
\begin{equation}
    \hat V_{\text{soc}}=\frac{\Omega\hbar}{2}\sum\limits_{j=1}^N \bigg(e^{\mathrm{i}\lambda d j}\, S_{j,+} + e^{-\mathrm{i}\lambda d j}\, S_{j,-}\bigg).
    \label{eq:V_soc}
\end{equation}
It can be represented in terms of the $q$-S-arrow operators:
\begin{equation}
   \hat V_{\text{soc}}=\frac{\hbar\Omega}{2}\sum\limits_{m=-s}^{s-1} \alpha_{s,m}\Big( S_q^{m\rightarrow m+1} + S_{-q}^{m+1\rightarrow m} \Big).
\end{equation}
This representation is particularly useful, as it explicitly shows that $\hat V_{\mathrm{soc}}$ directly couples the GPES and $q$-PES sectors. Therefore, the PES framework provides a natural basis for performing perturbative calculations with $\hat V_{\mathrm{soc}}$ as the perturbation.

The second-order Schrieffer--Wolff expression~\cite{BRAVYI20112793} for the effective Hamiltonian within the $E=0$ manifold generated by $\hat V_{\mathrm{soc}}$ is
\begin{equation}
    \hat{H}_{\rm GPES}^{(\mathrm{soc})}=\hat P(\text{\small GPES})\, \hat V_{\text{soc}}\sum_{q\neq 0}\frac{\hat P(\text{\small $q$-PES})}{-E_q}\, \hat V_{\text{soc}}\,\hat P(\text{\small GPES}),
    \label{eq:eff^soc_GPES}
\end{equation}
where $\hat P(\mathrm{GPES})$ projects onto the GPES subspace and $\hat P(q\text{-PES})$ onto the subspace of $q$-PESs with quasi-momentum $q$.

As shown in Eq.~\eqref{eq:q-PES}, the $q$-S-arrow operators generate $q$-PESs from GPESs according to
\begin{equation}
    |q,m_q;\vec{n}'\rangle = (\mathcal{Q}_{n_{m_q}}^{n_\mu})^{-1/2}\,\hat{S}^{\mu\rightarrow m_q}_q\,|\vec{n}\rangle,
    \label{GPES_to_qPES}
\end{equation}
while the inverse process, mapping a $q$-PES back to the GPES manifold, is obtained through
\begin{equation}
    (\mathcal{Q}_{n_{m_q}}^{n_\mu})^{-1/2}\,\hat{S}^{m_q\rightarrow\mu }_{-q}\,|q,m_q;\vec{n}'\rangle = |\vec{n}\rangle\label{qPES_to_GPES}
\end{equation}

The detailed derivation of the effective model generated by the light coupling is presented in Appendix~\ref{apx:deriv_light}.

When $\phi\neq\pi$ the effective model is
\begin{equation}
     \hat{H}^{({\mathrm{soc}},\phi)}_{\rm GPES}=\chi_\phi
     \Bigg(
     \hat J^2 - \hat{J}_z^2
     -
     \frac{N}{2}\sum_{m=-s}^{s-1}
     \alpha_{s,m}^2
     (\hat n_m+\hat n_{m+1})
     \Bigg),
    \label{eq:corr1}
\end{equation}
with
\begin{equation}
    \chi_\phi = \frac{\hbar^2\Omega^2}
     {4\mathrm{J_{SE}}(\cos \phi-1)(N-1)}.
\end{equation}
The population-dependent contribution (the last term) breaks the conservation of the total collective spin. For $s=\frac{1}{2}$, however, this term reduces to the conserved total atom number $N$, recovering the OAT model with the emergent $\hat J_z^2$ dynamics reported in Ref.~\cite{Tana2022}.

\begin{figure}
    \centering
    \includegraphics[width=0.9\linewidth]{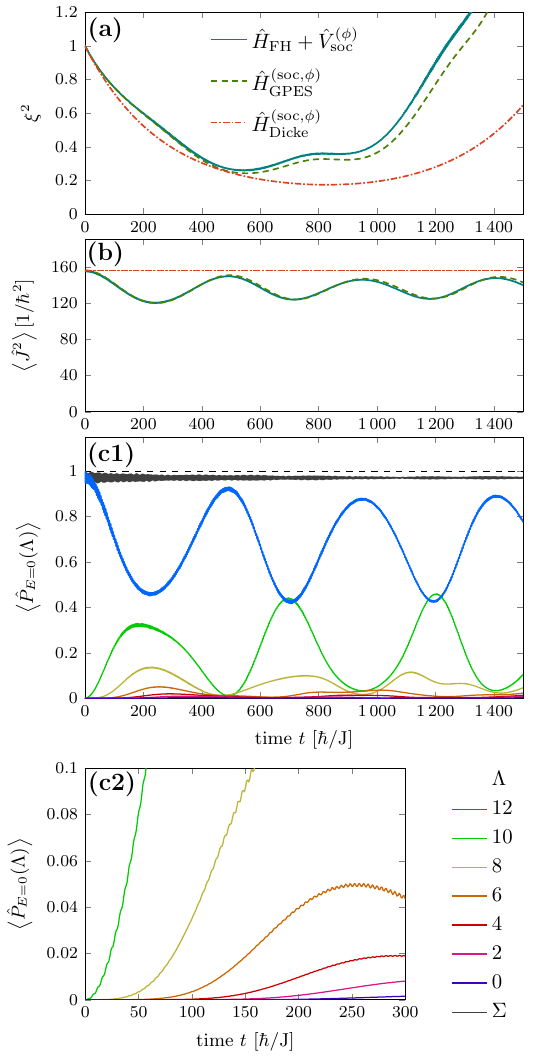}
    \caption{
    Spin dynamics and spin squeezing for $s=1$ fermions ($N=12$ sites) in the strongly interacting regime ($U=5{\rm J}$) with laser coupling ($\Omega=0.1{\rm J}$) at a phase $\phi=5\cdot 2\pi/N$. The initial coherent state is $|\vartheta{=}\frac{\pi}{2},\varphi{=}0\rangle=e^{\mathrm{i}\hat{J}_y\frac{\pi}{2}}|\mathrm{max}\rangle$ (pointing in $x$ direction).}
    \label{fig:plot_phi}
\end{figure}

It is important to note that for $s\geq1$, restricting the perturbative calculation to the Dicke and spin-wave subspaces only,
\begin{equation}
    \hat{H}_{\rm Dicke}^{(\mathrm{soc})}=\hat P(\text{\small Dicke})\, \hat V_{\text{soc}}\sum_{q\ne 0}\frac{\hat P(\text{\small SWS})}{-E_q}\, \hat V_{\text{soc}}\,\hat P(\text{\small Dicke})\label{eq:eff^soc_Dicke}
\end{equation}
also yields an OAT-like effective Hamiltonian (for $\phi\neq\pi$),
\begin{equation}
    \hat{H}_{\rm Dicke}^{(\mathrm{soc},\phi)}
    =\frac{\hbar^2\Omega^2}{4\mathrm{J}_{\rm SE}(1-\cos\phi)}\,
    \frac{1}{2Ns-1}\,\hat{J}_z^2.\label{eq:eff^soc_Dicke_phi}
\end{equation}
However, $\hat{H}_{\rm Dicke}^{(\mathrm{soc})}$ does not provide the correct effective description for $s\geq1$ as discussed in detail by us in ~\cite{short}, since it neglects additional states belonging to the $E=0$ and $E_q$ manifolds outside the Dicke and SWS subspaces. 
Because the initial state belongs to the Dicke manifold, this restricted model reproduces only the very initial stage of the dynamics. At later times, the evolution explores the full GPES manifold, and the restriction to Dicke states becomes invalid. The Dicke-based effective Hamiltonian is exact only for $s=1/2$, where the Dicke and SWS manifolds are complete for relevant energy sectors, and Eq.~\eqref{eq:corr1} is recovered.

Returning to the effective models derived within the PES framework, the special case $\phi=\pi$ gives
\begin{multline}
    \hat{H}^{({\mathrm{soc}},\pi)}_{\text{GPES}}=\chi_\pi
    \Bigg( 4\hat{J}_x^2-  N\!\sum\limits_{m=-s}^{s-1}  \alpha_{s,m}^2\Big(\hat{n}_m+\hat{n}_{m+1}\Big)-\\
    -  N\!\sum\limits_{m=-s}^{s-2}  \alpha_{s,m}\alpha_{s,m+1}\Big(\hat{S}^{m\rightarrow m+2} + \hat{S}^{m+2\rightarrow m}\Big)\Bigg),
    \label{eq:corr2}
\end{multline}
where
\begin{equation}
    \chi_\pi=\frac{\hbar^2\Omega^2}{-8\mathrm{J_{SE}}(N-1)}.
\end{equation}
In the spin-$1/2$ case, the last term vanishes identically because no transitions between magnetic sub-levels separated by two units are possible, while the population-dependent term reduces to a constant. Consequently, the effective Hamiltonian reduces to the OAT form with dynamics governed by $\hat J_x^2$.

Similarly, for $\phi=\pi$, an incorrect restriction of the perturbative calculation to the Dicke and SWS subspaces in Eq.~\eqref{eq:eff^soc_Dicke}, also yields an OAT Hamiltonian:
\begin{equation}
    \hat{H}_{\rm Dicke}^{(\mathrm{soc},\phi)}
    =-\frac{\hbar^2\Omega^2}{2\mathrm{J}_{\rm SE}}\,
    \frac{1}{2Ns-1}\,\hat{J}_x^2,
    \label{eq:eff^soc_Dicke_pi}
\end{equation}
and this restricted description is exact only for $s=\frac{1}{2}$.

\begin{figure}
    \centering
    \includegraphics[width=0.9\linewidth]{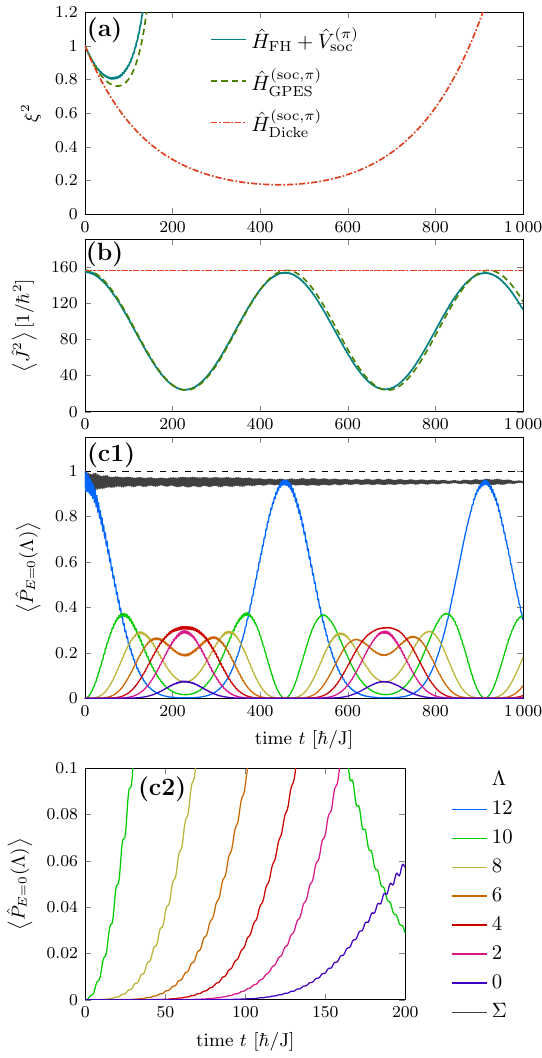}
    \caption{Spin dynamics and squeezing for $s=1$ fermions ($N=12$ sites) in the strongly interacting regime ($U=5\mathrm{J}$) with laser coupling ($\Omega=0.1\mathrm{J}$) at a specific phase $\phi=6\cdot\frac{2\pi}{N}$. The initial coherent state is $|\vartheta{=}0,\varphi{=}0\rangle=|\mathrm{max}\rangle$ (pointing in $z$ direction).
    }
    \label{fig:plot_pi}
\end{figure}


Figures~\ref{fig:plot_phi} and \ref{fig:plot_pi} present exact unitary evolution results for $s=1$ fermions in $N=12$ lattice sites in the strongly interacting regime ($U=5\mathrm{J}$), weakly coupled to a laser field ($\Omega=0.1\mathrm{J}$). The two figures correspond to different spin-orbit-coupling phases, $\phi=5\cdot 2\pi/N$ and $\phi=\pi$, respectively. In both cases, panel (a) compares the squeezing parameter $\xi^2$ obtained from the full Fermi--Hubbard model with to light coupling, $\hat{H}_{\rm FH}+\hat{V}_{\rm soc}$, with the predictions of the effective ground-population-eigenstate model, $\hat{H}_{\rm GPES}^{(\rm soc)}$, and the restricted Dicke-manifold approximation, $\hat{H}_{\rm Dicke}^{(\rm soc)}$ (that was correct in $s=\frac{1}{2}$ case). The effective model $\hat{H}_{\rm GPES}^{(\rm soc)}$ reproduces the full many-body dynamics with high accuracy over the entire evolution (the curves become indistinguishable when deepens od perturbative regime is increased, see \emph{Applicability of utilized perturbation} section in End Matter of~\cite{short}), whereas the Dicke approximation exhibits substantial deviations (that do not disappear in deeper perturbative regime), demonstrating that the squeezing dynamics can not be described within a single collective-spin manifold.

The origin of this breakdown is revealed in panel (b), which shows the evolution of the collective spin length $\langle \hat{J}^2\rangle$. While the Dicke description predicts a conservation of total spin, the full and effective dynamics display pronounced oscillations of $\langle \hat{J}^2\rangle$, indicating coherent coupling between sectors of different collective-spin length within $E=0$ manifold. This mechanism is directly visualized in panels (c1) and (c2), which resolve the populations $\langle \hat{P}_{E=0}(\Lambda)\rangle$ of the collective-spin length sectors within the degenerate $E=0$ manifold. Starting from the fully polarized coherent state with $\Lambda=12$ (from the Dicke manifold), the population progressively spreads into lower-spin sectors, occupying values as low as $\Lambda=0$. The redistribution among different $\Lambda$ manifolds originates from the enhanced degeneracy and underlying $\mathrm{SU}(d)$ symmetry of the spin-exchange model for high-spin fermions, which allows the dynamics to explore multiple collective-spin sectors while remaining within the degenerate energy manifold. The qualitative agreement between the results obtained for $\phi=5\cdot 2\pi/N$ and $\phi=\pi$ demonstrates that this mechanism is robust with respect to the spin-orbit-coupling phase and constitutes a generic feature of squeezing dynamics in strongly interacting high-spin Fermi-Hubbard systems.

For $\phi=\pi$, the oscillations between different $\Lambda$ sectors are particularly pronounced and regular, resulting in large-amplitude oscillations of the collective-spin length. Importantly, even when the system returns completely to the $\Lambda=12$ Dicke manifold (e.g., around $t=450\,\hbar/\mathrm{J}$), the state is not restored to a spin-squeezed state and does not follow the squeezing dynamics predicted by the Dicke-restricted OAT model $\hat{H}_{\rm Dicke}^{(\mathrm{soc},\pi)}$.

The situation is qualitatively different for $\phi\neq\pi$. In this case, the $\Lambda$-sector populations oscillate predominantly among states with large collective spin, with only weak occupation of low-$\Lambda$ sectors. This leads to moderate oscillations of $\langle\hat J^2\rangle$, corresponding to a small reduction of the collective-spin length relative to the Dicke manifold. Furthermore, around $t=500\,\hbar/\mathrm{J}$, when the system returns predominantly to the Dicke manifold, the coherences responsible for squeezing are largely preserved during the spin-length oscillations, resulting in a squeezing parameter close to that predicted by the Dicke-restricted OAT model. In fact, the squeezing is enhanced when the collective-spin length reaches its maximum value, since the same reduced variance corresponds to a larger metrological gain. These results indicate a possible route toward generating useful spin squeezing in high-spin systems.

\section{Summary and Conclusions}

In this work, we have investigated the $\mathrm{SU}(d)$ spin-exchange model describing strongly interacting ultra-cold fermions with large nuclear spin ($s\ge 1$) in optical lattices. Our primary result is the systematic construction of the eigenstates structure of the spin-exchange Hamiltonian in terms of PES within a two-energy manifold. Exploiting the conservation of the collective population operators, we identified the GPES spanning the degenerate zero-energy manifold and the $q$-PES forming the neighboring energy quasi-momentum manifolds. This construction provides a natural basis for the description of high-spin fermionic systems, directly connected to experimentally accessible spin-population observables.

The derived PES framework reveals qualitative differences between spin-$1/2$ and higher-spin fermionic ultra-cold atoms weakly coupled to light. Whereas the zero energy dynamics of spin-$1/2$ models remain confined to the maximal-spin Dicke manifold, the extensive degeneracies associated with the underlying $\mathrm{SU}(d)$ symmetry allow systems with $s\ge1$ to dynamically explore multiple collective-spin sectors, since for given energy $E$ a broad range of spin length values $\Lambda$ is available. As a consequence, the collective spin length is generally not conserved, and descriptions restricted to a fixed-spin manifold fail to capture essential aspects of the many-body dynamics.

Our PES framework open several concrete directions for future work: the extension to inhomogeneous systems or non-unit filling where the singlon constraint is relaxed; the derivation of effective Hamiltonians for other light-matter coupling geometries beyond spin-orbit coupling; and the design of metrological protocols that deliberately exploit inter-manifold dynamics to achieve sensitivity beyond what is possible within a fixed collective-spin manifold.

\section*{ACKNOWLEDGMENTS}
The Authors gratefully acknowledge discussions with B.\ Laburthe-Tolra and M.\ Robert-de-Saint-Vincent. 
E.W.\ acknowledges the kind hospitality of the Laboratoire de Physique des Lasers.
This work was supported by the Polish National Science Center SHENG project DEC-2023/48/Q/ST2/00087.

\appendix

\section{S-arrows properties \label{apx:S-arrow}}

\subsection{On-site operators}

Definition:
\begin{equation}
    \hat S_j^{m\rightarrow m'} = \hat a_{j,m'}^\dagger \hat a_{j,m}
\end{equation}
Hermitian conjugation:
\begin{equation}
    \Big(\hat S_j^{m\rightarrow m'}\Big)^\dagger=\hat S_j^{m'\rightarrow m}
\end{equation}
Composition of S-arrows -- transitivity:
\begin{equation}
\hat S_j^{\tilde m'\rightarrow m''}\hat S_j^{m\rightarrow m'} = \hat S_j^{m\rightarrow m''}\,\delta_{\tilde{m}',m'}\label{S-arrow_trans}
\end{equation}
Commutation relation:
\begin{equation}
 \Big[ \hat S_j^{m\rightarrow m'}; \hat S_{\tilde{j}}^{\tilde{m}\rightarrow\tilde{m}'} \Big] = \delta_{j,\tilde{j}} \Big( \hat S_j^{\tilde{m}\rightarrow m'}\ \delta_{m,\tilde{m}'} - \hat S_j^{m\rightarrow\tilde{m}'}\ \delta_{m',\tilde{m}}\Big)\label{S-arrow-comm}
\end{equation}

\subsection{Collective operators}

Definition:
\begin{equation}
    \hat S^{m\rightarrow m'}=\sum\limits_{j=1}^N \hat S_j^{m\rightarrow m'}
\end{equation}
Hermitian conjugation:
\begin{equation}
    \Big(\hat S^{m\rightarrow m'}\Big)^\dagger=\hat S^{m'\rightarrow m}
\end{equation}
Commutation relation:
\begin{equation}
    \Big[ \hat S^{m\rightarrow m'}; \hat S^{\tilde{m}\rightarrow\tilde{m}'} \Big] = \hat S^{\tilde{m}\rightarrow m'}\ \delta_{m,\tilde{m}'} - \hat S^{m\rightarrow\tilde{m}'}\ \delta_{m',\tilde{m}}\label{eq:S-qrrow_col_kom}
\end{equation}
We define $\mathrm{SU(2)}$-subspace spin-z operator:
\begin{equation}
    \hat S_z^{m/m'}=\hat n_m-\hat n_{m'}
\end{equation}
It is a useful object, because:
\begin{equation}
    \Big[\hat S^{m'\rightarrow m}; \hat S^{m\rightarrow m'} \Big] = \hat S_z^{m/m'}
\end{equation}
$\hat S_z^{m/m'}$ are by construction hermitian:
\begin{equation}
    \Big(\hat S_z^{m/m'}\Big)^\dagger = \hat S_z^{m/m'}
\end{equation}
and antisymmetric with respect to $m\leftrightarrow m'$ exchange:
\begin{equation}
    \hat S_z^{m/m'} = -\hat S_z^{m'/m}
\end{equation}
Commutator of s-arrow and $\mathrm{SU(2)}$-subspace spin-z:
\begin{align}
    \Big[ \hat S^{m\rightarrow m'}; \hat S_z^{m/m'} \Big] &= 2 \hat S^{m\rightarrow m'}\\[3pt]
    \Big[ \hat S^{m'\rightarrow m}; \hat S_z^{m/m'} \Big] &= -2 \hat S^{m'\rightarrow m}
\end{align}

\section{Derivation of Spin Exchange Hamiltonian}

The Spin Exchange (SE) model $\hat H_{\mathrm{SE}}$ is obtained as an effective model $\hat H_{\mathrm{sing}}^{(\mathrm{t})}$ for Fermi-Hubbard system \eqref{eq:HFH} in $U\gg\mathrm{J}$ regime for states with one atom in each lattice site (singlons) with tunneling term of F-H $\hat H_{\mathrm{t}}$ taken as perturbation on top of interaction term $\hat H_{\mathrm{int}}$. 

The second order perturbative expression that leads to the SE model is given in \eqref{eq:pert_FH_to_SE}, when expressions for $\hat H_{\mathrm{t}}$ are plugged they lead to the emergence of two distinct gropus of terms:
\begin{multline}
    \hat H_{\mathrm{sing}}^{(\mathrm{t})}=-\frac{\mathrm{J}^2}{U}\sum_j\sum_{m,m'=-s}^{s}\cdot\\ \cdot\bigg(\underbrace{a_{j,m}^\dagger \hat a_{j+1,m}\hat a_{j+1,m'}^\dagger \hat a_{j,m'}}_{\mathrm{RL}}+ \underbrace{\hat a_{j+1,m}^\dagger \hat a_{j,m}\hat a_{j,m'}^\dagger \hat a_{j+1,m'}}_{\mathrm{LR}}\bigg)\label{eq:SE_in_aa}
\end{multline}
The  RL terms correspond to the processes where, between two sites tunneling happens first from the left one to the right and then from the right to the left. Analogously, LR terms describe processes with tunneling first from right to left and then from left to right.

Now the operators in all terms should be reordered so that the operators referring to the same site are gathered together. In this process, anticommutation relation for ferminonic creation-annihilation operators has to be taken in to account: $\big\{\hat a_{j,m}\hat a_{j',m'}^\dagger\big\} = \hat\delta_{j,j'} \hat\delta_{m,m'}$, $\big\{\hat a_{j,m}\hat a_{j',m'}\big\} =  \big\{\hat a^\dagger_{j,m}\hat a_{j',m'}^\dagger\big\} = 0$. Because of that, we treat terms with $m=m'$ and with $m\neq m'$, splitting the Hamiltonian into two parts:
\begin{equation}
    \hat H_{\mathrm{sing}}^{(\mathrm{t})}=\hat H_{\text{diag}} + \hat H_{\text{off-d}}
\end{equation}
\begin{align}
    \hat H_{\text{diag}}=&-\frac{\mathrm{J}^2}{U}\sum_j\sum_{m=-s}^{s} \Big(\hat a_{j,m}^\dagger \hat a_{j+1,m}\hat a_{j+1,m}^\dagger \hat a_{j,m}+\nonumber\\
    &\hspace{60pt}+ \hat a_{j+1,m}^\dagger \hat a_{j,m}\hat a_{j,m}^\dagger \hat a_{j+1,m}\Big)\label{eq:SE_diag}\\[5pt]
    \hat H_{\text{off-d}}=&-\frac{\mathrm{J}^2}{U}\sum_j\sum_{m\neq m'}\Big(\hat a_{j,m}^\dagger \hat a_{j+1,m}a_{j+1,m'}^\dagger \hat a_{j,m'}+ \nonumber\\
    &\hspace{60pt}+\hat a_{j+1,m}^\dagger \hat a_{j,m}\hat a_{j,m'}^\dagger \hat a_{j+1,m'}\Big)
    \label{eq:SE_off-d}
\end{align}
where we use notation $\sum\limits_{m\neq m'}=
\sum\limits_{m=-s}^s\Big(
\sum\limits_{m'=-s}^{m-1}+\!
\sum\limits_{m'=m+1}^s\Big)$.

For $m=m'$ terms commutation of creation and annihilation operators of the same site and magnetization leads to emergence of additional identity operator $\hat{\mathbb{I}}$ and grouped operators reproduce particle number operators $\hat{a}_{j,m}^\dagger\hat{a}_{j,m}=\hat{n}_{j,m}$ leading to:
\begin{multline}
    \hat a_{j,m}^\dagger \hat a_{j+1,m}\hat a_{j+1,m}^\dagger \hat a_{j,m}+ \hat a_{j+1,m}^\dagger \hat a_{j,m}\hat a_{j,m}^\dagger \hat a_{j+1,m}=\\[5pt]
    =\hat n_{j,m}(-\hat n_{j+1,m}+\hat{\mathbb{I}}) + \hat n_{j+1,m} \Big(-\hat n_{j,m}+\hat{\mathbb{I}}\Big)\label{eq:diag}
\end{multline}
where RL and LR terms were merged by changing dummy indices. Since for singlons we have $1=\sum_{m,=-s}^s n_{j,m,}$ for any $j$, we can further rewrite $m=m'$ terms using:
\begin{equation}
    -\hat n_{j,m}+\hat{\mathbb{I}} = \Bigg(
\sum\limits_{m'=-s}^{m-1}+\!
\sum\limits_{m'=m+1}^s\Bigg)\hat{n}_{j,m'}\label{eq:diag2}
\end{equation}
Applying \eqref{eq:diag} modified according to \eqref{eq:diag2} to \eqref{eq:SE_diag} and changing dummy indices to merge terms of RL and LR origin, we obtain $\hat H_{\text{diag}}$ in the form:
\begin{equation}
    \hat H_{\text{diag}}=-\frac{2\mathrm{J}^2}{U}
\sum\limits_{j=1}^{N}
\sum\limits_{m\neq m'}\hat{n}_{j,m} \hat{n}_{j+1,m'}
\end{equation}
We see that $\hat H_{\text{diag}}$ has a form of nearest neighbour interactions independent of magnetization, which, however do not occur if magnetizations are equal (what comes from Pauli exclusion).

For $m\neq m'$ terms, commuting creation-annihilation operators gives only several cancelling minus signs leading to:
\begin{multline}
    \hat a_{j,m}^\dagger \hat a_{j+1,m}\hat a_{j+1,m'}^\dagger \hat a_{j,m'}+ \hat a_{j+1,m}^\dagger \hat a_{j,m}\hat a_{j,m'}^\dagger \hat a_{j+1,m'}=\\[5pt]
    =-\hat S^{m'\rightarrow m}_j \hat S^{m\rightarrow m'}_{j+1} - \hat S^{m\rightarrow m'}_{j+1}  \hat S^{m'\rightarrow m}_j \label{eq:off-d}
\end{multline}
Plugging \eqref{eq:off-d} in to \eqref{eq:SE_off-d} and changing dummy indices to merge together terms of RL and LR origin, we obtain the  $\hat H_{\text{off-d}}$:
\begin{equation}
     \hat H_{\text{off-d}}=\frac{2\mathrm{J}^2}{U}
\sum\limits_{j=1}^{N}
\sum\limits_{m\neq m'}\hat{S}^{m'\rightarrow m}_j \hat{S}^{m\rightarrow m'}_{j +1}
\end{equation}
We see that $\hat H_{\text{off-d}}$ is responsible for nearest neighbor spin exchange effective processes with strength independent of magnetization, , however do not occur if magnetizations are equal (what comes from Pauli exclusion).

An effective model obtained in this way is called the Spin Exchange effective Hamiltonian:
\begin{equation}
   \hat H_{\mathrm{sing}}^{(\mathrm{t})} =
\mathrm{J_{SE}}
\sum\limits_{j=1}^{N}
\sum\limits_{m\neq m'} 
\Big(
\hat{S}^{m'\rightarrow m}_j \hat{S}^{m\rightarrow m'}_{j +1}
-\hat{n}_{j,m} \hat{n}_{j+1,m'}
\Big) 
\end{equation}
denoted later by $\hat{H}_{\mathrm{SE}}$ (with $\mathrm{J_{SE}}=\frac{2\mathrm{J}^2}{U}$).

\section{Alternative forms of SE Hamiltonian \label{apx:alt_SE}}

The default for this work form of Spin Exchange Hamiltonian is given by \eqref{eq:HSE}. Here we present several equivalent forms of $\hat{H}_\mathrm{SE}$ that are used in literature. 

Since $\hat n_{j,m}=\hat a^\dagger_{j,m}\hat a_{j,m}=\hat S^{m\rightarrow m}_j$:
\begin{equation}
\hat{H}_\mathrm{SE} = \frac{2\mathrm{J}^2}{U}\sum\limits_{j=1}^{M'}
\sum\limits_{m\neq m'} \Big(-\hat S^{m\rightarrow m}_j \hat S^{m'\rightarrow m'}_{j +1}+ \hat S^{m'\rightarrow m}_j \hat S^{m\rightarrow m'}_{j +1}\Big).
\end{equation}
This form could be useful if one would like to work with the algebra of S-arrow operators (instead of creation-annihilation).

One can notice that:
\begin{align}
&\Big[-\hat S^{m\rightarrow m}_j \hat S^{m'\rightarrow m'}_{j +1}+ \hat S^{m'\rightarrow m}_j \hat S^{m\rightarrow m'}_{j +1}\Big]_{m'=m} \nonumber \\
&= -\hat S^{m\rightarrow m}_j \hat S^{m\rightarrow m}_{j +1}+ \hat S^{m\rightarrow m}_j \hat S^{m\rightarrow m}_{j +1} = 0
\end{align}
As such a term is identically zero, we can add it to $\hat{H}_\mathrm{SE}$ in this way, getting rid of the exclusion $m\neq m'$ in the summation:
\begin{equation}
\hat{H}_\mathrm{SE} = \frac{2\mathrm{J}^2}{U}\sum\limits_{j=1}^{M'}
\ \sum\limits_{m, m'=-s}^{s} \!\!\!\Big(\hat S^{m'\rightarrow m}_j \hat S^{m\rightarrow m'}_{j +1}\!{-}\hat S^{m\rightarrow m}_j \hat S^{m'\rightarrow m'}_{j +1}\Big)
\end{equation}
or after restoring particle number operators in notation:
\begin{equation}
\hat{H}_\mathrm{SE} = \frac{2\mathrm{J}^2}{U}\sum\limits_{j=1}^{M'}
\ \sum\limits_{m, m'=-s}^{s}\!\! \Big(\hat S^{m'\rightarrow m}_j \hat S^{m\rightarrow m'}_{j +1}{-}\hat n_{j,m}\hat  n_{j+1,m'}\Big)\label{H^SE_no_m_neq_m'}
\end{equation}

In the literature, one can also see another common way of expressing the spin exchange Hamiltonian, which looks simpler, however, turns out to be less practical for our application. It is obtained if \eqref{eq:diag} is left as it is, without applying \eqref{eq:diag2}. Then $\hat{H}_\text{diag} $ reads:
\begin{align}
\hat{H}_\text{diag} &= 
-\frac{\mathrm{J}^2}{U}\sum\limits_{j}
\sum\limits_{m=-s}^{s} 
\Big(
\hat n_{j,m} 
\big(- \hat n_{j+1,m} 
+ \hat{\mathbb{I}}\big)
 \nonumber \\[-5pt]
&
\hspace{100pt}+ \hat n_{j+1,m} \big(-\hat n_{j,m}+\hat{\mathbb{I}}\big) \Big) =\nonumber\\
&= \frac{\mathrm{J}^2}{U}\sum\limits_{j}\sum\limits_{m=-s}^{s} \bigg(2\, \hat n_{j,m}  \hat n_{j+1,m} 
-\hat n_{j,m} - \hat n_{j+1,m}\bigg)
\end{align}
For singlons states (with one atom in each lattice site) we have:
\begin{equation}
\sum\limits_{j=1}^N\sum\limits_{m=-s}^{s} n_{j,m} = N
\end{equation}
since $\sum\limits_{m=-s}^{s} n_{j,m}=1$. What leads to $\hat{H}_\text{diag} $:
\begin{equation}
\hat{H}_\text{diag} = \frac{2\mathrm{J}^2}{U}\sum\limits_{j}\sum\limits_{m=-s}^{s} \hat n_{j,m} \hat n_{j+1,m} 
- \frac{2\mathrm{J}^2}{U}\, N
\end{equation}
or with the use of S-arrow operators:
\begin{equation}
\hat{H}_\text{diag}= \frac{2\mathrm{J}^2}{U}\sum\limits_{j}\sum\limits_{m=-s}^{s} \hat S_{j}^{m\rightarrow m} \hat S_{j+1}^{m\rightarrow m} 
- \frac{2\mathrm{J}^2}{U}\, N
\end{equation}
Then the total SE Hamiltonian reads:
\begin{equation}
\hat{H}_\text{SE} = \frac{2\mathrm{J}^2}{U}\Bigg[\sum\limits_{j}\,
\sum\limits_{m,m'=-s}^{s} \hat S^{m'\rightarrow m}_j \hat S^{m\rightarrow m'}_{j +1} - N\Bigg]\label{H^SE_Ray-N}
\end{equation}
where terms in $\hat{H}_\text{diag}$ and $\hat{H}_\text{off-d}$ were complementary to restore summation over whole ranges of $m$ and $m'$. The global change of energy offset by $+\frac{2\mathrm{J}^2}{U}N$ can be applied simplifying the SE Hamiltonian further:
\begin{equation}
\hat{H}_\text{SE} = \frac{2\mathrm{J}^2}{U}\sum\limits_{j}\,
\sum\limits_{m,m'=-s}^{s} \hat S^{m'\rightarrow m}_j \hat S^{m\rightarrow m'}_{j +1}\label{H^SE_Ray}
\end{equation}

The \eqref{H^SE_Ray} form of SE Hamiltonian, although compact, is not practical for calculations -- for example, for states with all atoms occupying the same magnetic sub-level, the \eqref{eq:HSE} clearly gives zero, which is a direct consequence of all dynamics being frozen by Pauli exclusion. Drawing the same conclusion from \eqref{H^SE_Ray} form of the SE Hamiltonian is not so trivial.

The form favoured by us of SE Hamiltonian \eqref{eq:HSE} also nicely exposes the physics of underlying physical processes that can happen between nearest neighbors giving a good intuition for the investigation of the relevant physics of the model.

\begin{widetext}
\section{Detailed derivation of effective dynamics $\hat{H}_{\rm GPES}$ due to light coupling $\hat{V}_{\mathrm{soc}}$ \label{apx:deriv_light}}

If we write explicitly interactions $\hat V_{\text{soc}}$ in expression for $\hat{H}_{\rm GPES}$ \eqref{eq:eff^soc_GPES} we get:
\begin{equation}
    \hat{H}_{\rm GPES}=\frac{\hbar^2\Omega^2}{4} \hat P(\text{\small GPES})\! \sum\limits_{m=-s}^{s-1}\! \alpha_{s,m}\Big( \hat S_q^{m\rightarrow m+1} + \hat S_{-q}^{m+1\rightarrow m} \Big)\sum_{\tilde q}\frac{\hat P(\text{\small $\tilde q$-PES})}{-E_{\tilde q}}\! \sum\limits_{m'=-s}^{s-1}\! \alpha_{s,m'}\Big( \hat S_q^{m'\rightarrow m'+1} + \hat S_{-q}^{m'+1\rightarrow m'} \Big) \hat P(\text{\small GPES})\label{eq:eff_VV}
\end{equation}
Let us see the effect of this Hamiltonian step by step: One starts from GPES state $\hat P(\text{\small GPES})$. Then by action of $\hat S_q^{m'\rightarrow m'+1}$ or $\hat S_{-q}^{m'+1\rightarrow m'}$ a GPES state is brought to respective $q$-PES or $-q$-PES state, both options with the same energy $E_q=E_{-q}=(4\mathrm{J}^2/U)(\cos q\lambda -1)$ (due to symmetry of cosine). In this way, from the sum over $\tilde q$ only terms with $\tilde{q}=\pm q$ are meaningful -- both give the same energy factor $E_q$. Next, for $q$-PES generated by $\hat S_q^{m'\rightarrow m'+1}$, only $\hat S_{-q}^{m+1\rightarrow m}$ can bring the state back to GPES manifold (according to equations \eqref{GPES_to_qPES} and \eqref{qPES_to_GPES}). Similarly, for $q$-PES generated $\hat S_{-q}^{m'+1\rightarrow m'}$ by only $\hat S_q^{m\rightarrow m+1}$ can bring state back to GPES manifold. On the other hand, for $q\neq\pi/d$ terms $\hat S_q^{m\rightarrow m+1}\hat S_q^{m'\rightarrow m'+1}$ and $\hat S_{-q}^{m+1\rightarrow m}\hat S_{-q}^{m'+1\rightarrow m'}$ won't give any contribution when sandwiched between projectors on GPES space $\hat P(\text{\small GPES})$. In this way only terms $\hat S_{-q}^{m+1\rightarrow m}\hat S_q^{m'\rightarrow m'+1}$ and $\hat S_q^{m\rightarrow m+1}\hat S_{-q}^{m'+1\rightarrow m'}$ plays non-trivial role in $\hat{H}^{(\phi)}_{\rm GPES}$ Hamiltonian, since they bridge GPES states to GPES states via $q$-PES states.

Putting these facts all together, we can write $\hat{H}^{(\phi)}_{\rm GPES}$ in $\phi\neq\pi$ as:
\begin{equation}
    \hat{H}^{(\phi)}_{\rm GPES}=\frac{\hbar^2\Omega^2}{-4E_q} \sum\limits_{m,m'=-s}^{s-1}\!\! \alpha_{s,m}\alpha_{s,m'}
    \Big(\hat S_q^{m\rightarrow m+1}\hat S_{-q}^{m'+1\rightarrow m'} + \hat S_{-q}^{m+1\rightarrow m}\hat S_q^{m'\rightarrow m'+1}\Big) \hat P(\text{\small GPES})
\end{equation}
To calculate this effective Hamiltonian, it is essential to find how these double $q$-S-arrow terms act on general GPES $|\vec{n}\rangle$ -- it turns out to be:
\begin{align}
     \hat S_{-q}^{m+1\rightarrow m}\hat S_q^{m'\rightarrow m'+1}\,|\vec{n}\rangle &= \delta_{m,m'}\, \frac{n_{m}(N-n_{m+1}-1)}{N-1}\, |\vec{n}\rangle -\nonumber\\[5pt]
     &\hspace{20pt}-\tilde\delta_{m,m'}\, \frac{1}{N-1}\sqrt{n_{m'}(n_{m'+1}{+}1)(n_{m+1}{-}\delta_{m+1,m'})(n_m{+}1{+}\delta_{m,m'+1})}\,\big|\vec{n}_{m',\underline{m}}\big\rangle\\[10pt]
     \hat S_q^{m\rightarrow m+1}\hat S_{-q}^{m'+1\rightarrow m'}\,|\vec{n}\rangle &= \delta_{m,m'}\, \frac{n_{m+1}(N-n_{m}-1)}{N-1}\, |\vec{n}\rangle -\nonumber\\[5pt]
     &\hspace{20pt}-\tilde\delta_{m,m'}\, \frac{1}{N-1}\sqrt{n_{m'+1}(n_{m'}{+}1)(n_m{-}\delta_{m,m'+1})(n_{m+1}{+}1{+}\delta_{m+1,m'})}\,\big|\vec{n}_{\underline{m'},{m}}\big\rangle
\end{align}
Where $\tilde\delta_{m,m'}=1-\delta_{m,m'}$ and $\vec{n}_{m',\underline{m}}=\big(n_{-s},\ldots,n_{m'}{-}1,n_{m'+1}{+}1,\ldots,n_{m}{+}1,n_{m+1}{-}1,\ldots,n_s\big)$ (explanation of index notation: not-underlined $m$ means one atom brought from $m$ population to $m+1$ and underlined $m$ means one atom brought from $m+1$ to $m$ magnetization). The above formulas comes directly from \eqref{eq:q-PES_non-ort} where non-orthogonality of $q$-PES with different $m_q$ plays crucial role.
The effective Hamiltonian $\hat{H}^{(\phi)}_{\rm GPES}$ can be written as:
\begin{align}
    \hat{H}^{(\phi)}_{\rm GPES}&=\frac{\hbar^2\Omega^2}{-4E_q(N-1)} \sum\limits_{m,m'=-s}^{s-1} \!\! \alpha_{s,m}\alpha_{s,m'} \!\!\sum\limits_{\vec{n}\in\{\text{GPES}\}}\nonumber\\
    &\hspace{20pt}\bigg[\delta_{m,m'}\Big(N(n_m+n_{m+1})-n_m(n_{m+1}+1)-n_{m+1}(n_m+1)\Big)\, |\vec{n}\rangle\langle\vec{n}|-\nonumber\\
    &\hspace{25pt}-\tilde\delta_{m,m'}\sqrt{n_{m'}(n_{m'+1}{+}1)(n_{m+1}{-}\delta_{m+1,m'})(n_m{+}1{+}\delta_{m,m'+1})}\,\big|\vec{n}_{m',\underline{m}}\big\rangle\langle\vec{n}|-\nonumber\\
    &\hspace{25pt}-\tilde\delta_{m,m'}\sqrt{n_{m'+1}(n_{m'}{+}1)(n_m{-}\delta_{m,m'+1})(n_{m+1}{+}1{+}\delta_{m+1,m'})}\,\big|\vec{n}_{\underline{m'},{m}}\big\rangle\langle\vec{n}|\ \bigg]
\end{align}

Since the action of the S-arrow has no GPES is:
\begin{equation}
    \hat S^{m\rightarrow \tilde m}\,|\vec{n}\rangle = \sqrt{n_m(n_{\tilde{m}}+\tilde\delta_{m,\tilde m})}\,\big|\vec{n}^{m\rightarrow \tilde m}\big\rangle, \label{S_arrow_on_GPES}
\end{equation}
(where vector $\vec{n}^{m\rightarrow \tilde m}$ has components related to the components of $\vec n$ is such a way: ${n}^{m\rightarrow \tilde m}_\nu=n_\nu-\delta_{\nu,m}+\delta_{\nu,\tilde m}$). The S-arrow operator in the GPES subspace can be written as:
\begin{equation}
    \hat S^{m\rightarrow \tilde m}P(\text{\small GPES})=\sum\limits_{\vec{n}\in\{\text{GPES}\}}\sqrt{n_m(n_{\tilde{m}}+\tilde\delta_{m,\tilde m})}\,\big|\vec{n}^{m\rightarrow \tilde m}\big\rangle\langle\vec{n}|
\end{equation}
What lets to express $\hat{H}^{(\phi)}_{\rm GPES}$ in the form:
\begin{align}
\hat{H}^{(\phi)}_{\rm GPES} &= \frac{\Omega^2\hbar^2}{-E_q 4(N-1)} \sum\limits_{m,m'=-s}^{s-1}\!\! \alpha_m \alpha_{m'}\ \cdot\nonumber\\
&\quad \cdot \bigg(\delta_{m,m'} \Big[ N \hat S^{m\to m} - \hat S^{m+1\to m}\hat S^{m\to m+1} + N \hat S^{m+1\to m+1} - \hat S^{m\to m+1} \hat S^{m+1\to m} \Big] -\nonumber\\
&\quad \phantom{\cdot} \phantom{\bigg(\bigg( } \tilde\delta_{m,m'} \Big[\hat S^{m+1\to m}\hat S^{m'\to m'+1} + \hat S^{m\to m+1} \hat S^{m'+1\to m'}\Big]\bigg)
\end{align}
where one can restore collective $\hat J_\pm$ operators and magnetization population numbers $\hat n_m$ simplifying the expression:
\begin{equation}
    \hat{H}^{(\phi)}_{\rm GPES}=\frac{\hbar^2\Omega^2}{4E_q(N-1)}\Bigg(\hat J_- \hat J_+ + \hat J_+ \hat J_- -  N\!\sum\limits_{m=-s}^{s-1}  \alpha_{s,m}^2\Big(\hat n_m+\hat n_{m+1}\Big)\Bigg)
\end{equation}
which is the effective Hamiltonian given in \eqref{eq:corr1} (since $\hat{J}^2=\hat{J}_-\hat{J}_+ + \hat{J}_+ \hat{J}_- + \hat{J}_z^2$).

The above consideration was for $q\neq\pi/\lambda$. In the specific case of $q=\pi/\lambda$ the left-propagating and right-propagating waves are indistinguishable, what we see by the fact that $q$-S-arrows for $q=-\pi/\lambda$ and $q=+\pi/\lambda$ are the same:
\begin{equation}
    \Sarrow{m}{m'}{q=+\pi}=\Sarrow{m}{m'}{q=-\pi}\equiv\Sarrow{m}{m'}{\pm\pi}
\end{equation}
Where, for convenience, we set $\lambda=1$.

Then it means that in \eqref{eq:eff_VV} terms $\hat S_q^{m\rightarrow m+1}\hat S_q^{m'\rightarrow m'+1}$ and $\hat S_{-q}^{m+1\rightarrow m}\hat S_{-q}^{m'+1\rightarrow m'}$ are non-trivial in GPES subspace, and the effective Hamiltonian has to be considered as:
\begin{multline}
    \hat{H}^{(\pi)}_{\text{GPES}}=\frac{\hbar^2\Omega^2}{-4E_q} \sum\limits_{m,m'=-s}^{s-1}\!\! \alpha_{s,m}\alpha_{s,m'}\,\cdot\\
    \cdot\Big(\hat S_{\pm\pi}^{m\rightarrow m+1}\hat S_{\pm\pi}^{m'\rightarrow m'+1} + \hat S_{\pm\pi}^{m\rightarrow m+1}\hat S_{\pm\pi}^{m'+1\rightarrow m'} + \hat S_{\pm\pi}^{m+1\rightarrow m}\hat S_{\pm\pi}^{m'\rightarrow m'+1} +  \hat S_{\pm\pi}^{m+1\rightarrow m}\hat S_{\pm\pi}^{m'+1\rightarrow m'}\Big) \hat P(\text{\small GPES})
\end{multline}
The action of $\hat S_{\pm\pi}^{m\rightarrow m+1}\hat S_{\pm\pi}^{m'+1\rightarrow m'}$ and $\hat S_{\pm\pi}^{m+1\rightarrow m}\hat S_{\pm\pi}^{m'\rightarrow m'+1}$ on GPES is known from $q\neq\pi/d$ case. Only action of $\hat S_{\pm\pi}^{m\rightarrow m+1}\hat S_{\pm\pi}^{m'\rightarrow m'+1}$ and $\hat S_{\pm\pi}^{m+1\rightarrow m}\hat S_{\pm\pi}^{m'+1\rightarrow m'}$ on GPES remains to be calculated:
\begin{align}
     \hat S_{\pm\pi}^{m\rightarrow m+1}\hat S_{\pm\pi}^{m'\rightarrow m'+1}\,|\vec{n}\rangle &= \delta_{m,m'+1}\, \frac{N-n_{m'+1}-1}{N-1}\sqrt{n_{m'}(n_{m'+2}+1)}\, \big|\vec{n}_{m',m'+1}\big\rangle -\nonumber\\[5pt]
     &\hspace{20pt}-\tilde\delta_{m,m'+1}\, \frac{1}{N-1}\sqrt{n_{m'}(n_{m'+1}{+}1)(n_{m}{-}\delta_{m,m'})(n_{m+1}{+}1{+}\delta_{m,m'})}\,\big|\vec{n}_{m',m}\big\rangle=\nonumber\\[5pt]
     &=-\frac{1}{N-1}\Big(\hat S^{m'+1\rightarrow m'+2}\,\hat S^{m'\rightarrow m'+1}\Big)+ \delta_{m,m'+1}\frac{N}{N-1}\hat S^{m'\rightarrow m'+2}\\[20pt]
     \hat S_{\pm\pi}^{m+1\rightarrow m}\hat S_{\pm\pi}^{m'+1\rightarrow m'}\,|\vec{n}\rangle &= \delta_{m+1,m'}\, \frac{N-n_{m-}-1}{N-1}\sqrt{n_{m'+1}(n_{m'-1}+1)}\, \big|\vec{n}_{\underline{m'},\underline{m'-1}}\big\rangle -\nonumber\\[5pt]
     &\hspace{20pt}-\tilde\delta_{m+1,m'}\, \frac{1}{N-1}\sqrt{n_{m'+1}(n_{m'}{+}1)(n_{m+1}{-}\delta_{m,m'})(n_{m}{+}1{+}\delta_{m,m'})}\,\big|\vec{n}_{\underline{m'},\underline{m}}\big\rangle=\nonumber\\[5pt]
     &=-\frac{1}{N-1}\Big(\hat S^{m'\rightarrow m'-1}\,\hat S^{m'+1\rightarrow m'}\Big)+ \delta_{m+1,m'}\frac{N}{N-1}\hat S^{m'+1\rightarrow m'-1}
\end{align}
(Eq.~\eqref{eq:q-PES_non-ort} was involved).
What when plugged in to $\hat{H}^{(\pi)}_{\text{GPES}}$ gives:
\begin{multline}
    \hat{H}^{(\pi)}_{\text{GPES}}=\frac{\hbar^2\Omega^2}{4E_q(N-1)}\cdot\\
    \cdot\Bigg(\underbrace{\hat J_+ \hat J_+ + \hat J_- \hat J_+ + \hat J_+ \hat J_- + \hat J_- \hat J_-}_{\displaystyle 4\hat J_x^2} -  N\!\sum\limits_{m=-s}^{s-1}  \alpha_{s,m}^2\Big(\hat n_m+\hat n_{m+1}\Big) -  N\!\sum\limits_{m=-s}^{s-2}  \alpha_{s,m}\alpha_{s,m+1}\Big(\hat S^{m\rightarrow m+2} + \hat S^{m+2\rightarrow m}\Big)\Bigg)
\end{multline}
what is \eqref{eq:corr2} from the main text.

\end{widetext}

\bibliography{biblio}

\end{document}